\documentclass[preprint]{aastex62}
\usepackage{amsmath,amssymb}
\usepackage{textcomp}
\usepackage{booktabs}
\usepackage{bm}

\newcommand\ltsima{$\; \buildrel <\over\sim \;$}
\newcommand\simlt{\lower.5ex\hbox{\ltsima}}
\newcommand\gtsima{$\; \buildrel >\over\sim \;$}
\newcommand\simgt{\lower.5ex\hbox{\gtsima}}

\newcommand{\mathbold}[1]{\mbox{\boldmath $\bf#1$}}
\newcommand\piEbold{{\mathbold \pi_{\rm E}}}
\newcommand\piEboldobs{{\mathbold \pi_{\rm E,obs}}}

\newcommand\murelbold{{\mathbold \mu_{\rm rel}}}

\shorttitle{{\it Spitzer} Microlens Parallax Systematic Errors} 
\shortauthors{Koshimoto \& Bennett}

\begin{document}


\title{Evidence of Systematic Errors in {\it Spitzer} Microlens Parallax Measurements}

\author{Naoki Koshimoto}
\affiliation{Department of Astronomy, Graduate School of Science, The University of Tokyo, 7-3-1 Hongo, Bunkyo-ku, Tokyo 113-0033, Japan}
\affiliation{Laboratory for Exoplanets and Stellar Astrophysics, NASA/Goddard Space Flight Center, Greenbelt, MD 20771, USA}
\affiliation{Department of Astronomy, University of Maryland, College Park, MD 20742, USA}

\author{David P. Bennett}
\affiliation{Laboratory for Exoplanets and Stellar Astrophysics, NASA/Goddard Space Flight Center, Greenbelt, MD 20771, USA}
\affiliation{Department of Astronomy, University of Maryland, College Park, MD 20742, USA}

\begin{abstract}
The microlensing parallax campaign with the {\it Spitzer} space telescope aims to measure 
masses and distances of microlensing events seen towards the Galactic bulge, with a focus on 
planetary microlensing events. The hope is to measure how the distribution of planets depends on 
position within the Galaxy. In this paper, we compare 50 microlens parallax measurements from the 2015 
{\it Spitzer} campaign to three different Galactic models commonly used in microlensing analyses,
and we find that $\geq 74\,$\% of these events have microlensing parallax values higher than
the medians predicted by Galactic models. 
The Anderson-Darling tests indicate probabilities of
$p_{\rm AD} < 6.6 \times 10^{-5}$ for these three Galactic models, while the binomial probability of 
such a large fraction of large microlensing parallax values is $< 4.6\times 10^{-4}$.
Given that many {\it Spitzer} 
light curves show evidence of large correlated errors, we conclude that this discrepancy is probably due
to systematic errors in the {\it Spitzer} photometry. 
We find formally acceptable probabilities of $p_{\rm AD} > 0.05$ for subsamples of events with bright source 
stars ($I_{\rm S} \leq 17.75$) or {\it Spitzer} coverage of the light curve peak.
This indicates that the systematic errors have a more serious influence on faint events, especially when the 
light curve peak is not covered by {\it Spitzer}.
We find that multiplying an error bar renormalization factor of 2.2 by the reported error bars on 
the {\it Spitzer} microlensing parallax measurements provides reasonable agreement 
with all three Galactic models.
However, corrections to the uncertainties in the {\it Spitzer} photometry itself are a 
more effective way to address the systematic errors. 
\end{abstract}


\section{Introduction}
The gravitational microlensing method \citep{mao91} is sensitive to planetary systems
at any distance between the Sun and the Galactic center. While distant planets can also be detected
by the transit method, microlensing is probably the best method to measure the 
planet distribution in our galaxy. A study of the Galactic distribution of planets can 
reveal the history of planet formation in our galaxy and the mechanism of planet formation in the 
Galactic bulge, which has a much higher density of stars than the Solar neighborhood.
Thus far, \citet{pen16} attempt a comparison of the distribution of distances to planetary microlens
systems with expectations based on a galactic model.
One aspect of the microlensing method which makes such a statistical study difficult is that the
lens mass $M_L$ and distance $D_L$ are not uniquely determined for most microlensing events.

To directly measure the lens mass, $M_L$, and estimate the distance, $D_L$, both the angular Einstein 
radius, $\theta_{\rm E}$, and the microlens parallax $\pi_{\rm E}$ must be measured.
The angular Einstein radius, $\theta_{\rm E}$, is given by  
\begin{equation}
\theta_{\rm E} \equiv \sqrt{\kappa M_L \pi_{\rm rel}} \ , 
\label{eq-thetaE}
\end{equation}
where $\kappa = 8.144 ~{\rm mas} ~M_\odot ^{-1}$, $\pi_{\rm rel} = {\rm 1\,AU}(D_L^{-1} - D_S^{-1} )$ 
and $D_S$ is the distance to the source star, which is approximately 8~kpc. The microlens parallax,
$\pi_{\rm E}$, is given by
\begin{equation}
\pi_{\rm E} \equiv \frac{\pi_{\rm rel}}{\theta_{\rm E}} \ ,
\label{eq-piE}
\end{equation}
and the lens mass can be obtained by eliminating $\pi_{\rm rel}$ from equations~\ref{eq-thetaE} and
\ref{eq-piE} to yield
\begin{equation}
M_L = {\theta_{\rm E}\over \kappa \pi_{\rm E} } \ .
\label{eq-mass}
\end{equation}

The angular Einstein radius can be measured when the finite source effect is seen in the light curve
or when the lens-source separation is measured after the microlensing event
\citep{ben15,bat15,bha18}. Microlensing parallax, $\pi_{\rm E}$, has traditionally been measured via the
detection of the effects of the Earth's orbital motion in the 
light curve \citep{macho-par1,an-eros2000blg5,mur11}.
Because these two effects are only occasionally measured, the only light curve parameter
that constrains the lens mass and distance is Einstein radius crossing time,
\begin{equation}
t_{\rm E} \equiv \frac{\theta_{\rm E}}{\mu_{\rm rel}} \ , 
\end{equation}
where $\mu_{\rm rel}$ is the lens-source relative proper motion.
For planetary events, the angular Einstein radius $\theta_{\rm E}$ is commonly measured because planetary
events usually show finite source effects, but the orbital microlensing parallax effect is detected 
only when the event's Einstein radius crossing time is relatively long. As a result, 
only $\sim 20 \%$ planetary events have their mass and distance determined by the 
combination of $\theta_{\rm E}$ and $\pi_{\rm E}$ \citep[e.g.,][]{benet10, mur11}.
For events where the finite source effect and/or the orbital parallax effect were not measured, 
probability distributions for the lens mass and distance can be estimated with a
Bayesian analysis using the Galactic model as its prior probability distribution,
under the assumption that the planet hosting probability does not depend on the lens 
mass and distance \citep[e.g.,][]{bea06, ben14}. It is also possible to determine the mass and
distance of the lens system by combining high angular resolution follow-up observations with
adaptive optics (AO) or the {\it Hubble Space Telescope} (HST) and mass luminosity relations
\citep{bat15, ben15, kos17a, kos17b, bha18}. However, these observations must be taken several
years after the event to measure the lens-source separation, depending on the $\mu_{\rm rel}$ value.

One might think of a statistical study of events with measurements of orbital microlensing parallax effects
to determine the Galactic distribution of planets, but
unfortunately, orbital microlensing parallax is difficult to detect for systems more distant than 
$D_L \approx 4\,$kpc \citep{sumi16,ben18}. \citet{pen16} did attempt to compare the planetary 
occurrence rate as a function of $D_L$, but this attempt was plagued by an inhomogeneous sample,
incorrect parallax measurements \citep{han_ob130723}, and overly optimistic detection efficiency 
estimates.

A more serious attempt to measure the Galactic distribution of planetary systems has been 
made with the {\it Spitzer} microlensing campaign \citep{yee15, uda15}, which is a systematic program
to make $\pi_{\rm E}$ measurements of microlensing events identified by ground-based surveys
since 2014. This program makes use of the $\sim 1\,$AU separation between {\it Spitzer}  and the 
Earth, to measure $\pi_{\rm E}$ for a carefully selected sample of events.
\citet{zhu17} did a statistical analysis of the 2015 {\it Spitzer} campaign, and they estimated 
that $\sim 1/3$ of all planet detections from the {\it Spitzer} campaign should be 
located in the bulge if the planet distributions are the same in the bulge as in the disk.

However, there are correlated systematic errors in many of the {\it Spitzer} light 
curves \citep{pol16, zhu17}, and they can potentially affect the microlensing parallax measurements.
\citet{zhu17} also discuss this. In particular, they describe that prominent deviations from single lens model, 
caused by the unknown systematics, are seen in the {\it Spitzer} light curves for 
5 events out of their raw sample of 50 events (see section 5.1 of their paper). In a handful of events
the {\it Spitzer} microlensing parallax measurements are consistent with a ground based parallax measurements
\citep{uda15,pol16,han17,shi17,wan17}. Also, some previous studies conducted tests on much
smaller samples of published events with {\it Spitzer} parallax measurements that might possibly be reinterpreted
as tests of consistency between the measurements and the Galactic model. 
\citet{sha19} considered a non-statistical sample of 13 published microlensing events with measurements of
$\theta_{\rm E}$ and $\pi_{\rm E}$ from {\it Spitzer} that are considered to be secure. They then compare the Bayesian
predictions for the lens system mass and distance to the results from {\it Spitzer} $\pi_{\rm E}$ measurements.
\citet{zang19} consider a sample of 8 published single-lens events without taking into account
``detection efficiency, and possible selection or publication biases\rlap." So, both of these
samples consist of events that the  {\it Spitzer} team selected to publish because they were considered 
to be ``interesting", which results in publication bias. 
Most of the events in these samples have a bright source star or a good  {\it Spitzer} light curve coverage 
over the peak or a caustic crossing. So, they have much stronger signals in the {\it Spitzer} data than is typical.
The published events are also less likely to have the obvious systematic photometry errors than a systematically
selected statistical sample.
Therefore, these comparisons are not precise enough to provide a useful test of the
precision of typical {\it Spitzer} microlensing parallax measurements. There are 
many events for which
the {\it Spitzer} light curve data have poor coverage of both the magnified portion of the light curve and the
baseline. These events might well have large systematic errors in the $\pi_{\rm E}$ measurements
due to systematic errors in the {\it Spitzer} photometry. Also, three microlensing events have been 
interpreted as lens systems located in the Galactic disk that are orbiting perpendicular
to or in the opposite direction of disk rotation based on {\it Spitzer} data with poor light curve coverage
\citep{shv17,shv19,chu19}.
The prior probability for such orbits is quite low ($\simlt 10^{-3}$), so it seems quite possible that the
microlensing parallax signals for these events are spurious due to systematic {\it Spitzer} photometry
errors.

In this paper, we compare the measurements of $t_{\rm E}$ and $\pi_{\rm E}$ for the sample of 50 single 
lens events from the 2015 {\it Spitzer} data \citep{zhu17}, to 
the predicted distributions based on Galactic models. We consider three different Galactic models previously used for 
microlensing studies, including the one used by \citet{zhu17}. 
We use a Bayesian analysis to compute the posterior distribution for the parallax measurement 
$f_{\rm post} (\pi_{\rm E})$ for each event in the sample. We then compare the medians, $\widetilde{ \pi}_{\rm E, post}$,
of these $\pi_{\rm E}$ posteriors to the predictions of the three Galactic models, and use the binomial distribution
to show that the distribution of $\widetilde{ \pi}_{\rm E, post}$ values in inconsistent with the Galactic models.
In order to explore this comparison in more detail, 
we convert each posterior $f_{\rm post} (\pi_{\rm E})$ distribution into a distribution called the
posterior inverse percentile distribution, and we compare the 
cumulative density function (CDF) of inverse percentile to the theoretical distribution
using the Anderson-Darling (AD) test. This indicates the null hypothesis that the observed distribution follows 
the model is rejected at high significance, $p_{\rm AD} \leq 6.6 \times 10^{-5}$, for all three of the Galactic
models that we consider. We also conduct these same tests for several sub-samples and find 
formally acceptable probabilities of $p_{\rm AD} > 0.05$ for a sub-sample of 17 events with bright source 
stars of $I_{\rm S} < 17.75$ and for a different 20 events sub-sample
with the light curve peak covered by {\it Spitzer}. However, for sub-samples with fainter source stars
or without {\it Spitzer} light curve coverage of the peak or caustic crossing, the probability is dramatically smaller.
We interpret this as evidence that the measured $\piEboldobs$ are contaminated
by the systematic photometry errors in the  {\it Spitzer} data, especially for faint events and events without
{\it Spitzer} light curve peak coverage.

This paper is organized as follows. In Section \ref{sec-method} we explain our method, 
focusing the basic idea on how we compare observations with a model without calculating detection efficiencies.
We explain the three Galactic models that we employ \citep{zhu17, sum11, ben14} and focus on the
differences between these models in Section \ref{sec-model}.
In Section \ref{sec-sample}, we explain the \citet{zhu17} sample and how we derive the posterior $\pi_{\rm E}$ distribution for each event.
Section \ref{sec-test} presents the results of our statistical comparison between the posterior and the 
model predicted $\pi_{\rm E}$ distributions using the Anderson-Darling (AD)
statistics. We present the same statistical tests with modified Galactic models in Section \ref{sec-discrep},
and we show that reasonable Galactic model modifications cannot explain the {\it Spitzer} measurements.
In the same section, we also discuss other potential factors that might possibly affect the results, 
and we show the discrepancy cannot be explained by these factors.
We discuss {\it Spitzer} systematic photometry errors in Section \ref{sec-dis}, and we describe our findings of correlations between the level of systematic errors in the $\pi_{\rm E}$ distribution and
the source brightness and {\it Spitzer} signal strength.
We present our conclusions in Section \ref{sec-con}.

\section{Method} \label{sec-method}
One of general difficulties involved in the comparison of an observational data set with a model is the determination
of detection efficiencies (or selection effects). The detection efficiency is defined as the probability that a 
microlensing event is selected to be part of the sample being studied.
The seven parameters that characterize a single lens are:  
the time of closest angular approach between the source and lens stars, $t_0$,
the impact parameter of the source trajectory with respect to the lens star, $u_0$, 
the Einstein radius crossing time, $t_{\rm E}$,
the angular Einstein radius, $\theta_{\rm E}$, the microlens parallax vector, 
${\bm \pi_{\rm E}}= \pi_{\rm E} \, {\bm \mu_{\rm rel}}/\mu_{\rm rel} =  (\pi_{\rm E,N}, \pi_{\rm E,E})$,
and the source flux $F_{\rm S}$. The microlensing parallax vector,
${\bm \pi_{\rm E}}$, is a vector with a magnitude of $\pi_{\rm E}$ and a direction parallel to
${\bm \mu_{\rm rel}}$. The north and east components are $\pi_{\rm E,N}$ and $\pi_{\rm E,E}$, 
respectively \citep{gou92}. Four of these parameters affect the detection efficiency of an event; 
$t_0$ determines the coverage of the light curve; $u_0$ determines the peak magnification;
$t_{\rm E}$ is the event duration, and $F_{\rm S}$ controls the brightness and photometric 
signal to noise ratio.
The parameters that provide information about the lens mass and the distance to the lens are 
$t_{\rm E}$, $\theta_{\rm E}$ and $\pi_{\rm E}$.
Therefore, the probability (density) for a single lens event to occur, be discovered, and then be
selected to be part of the sample being studied can be decomposed into three other functions,
\begin{eqnarray}
f_{\rm obs} (t_{\rm E}, \theta_{\rm E}, {\bm \pi_{\rm E}}, t_0, u_0, F_{\rm S}) \propto \Gamma_{\rm Gal} (t_{\rm E}, \theta_{\rm E}, {\bm \pi_{\rm E}}) \, \eta (t_0, u_0, F_{\rm S}) \, \epsilon (t_{\rm E}, t_0, u_0, F_{\rm S}) \ ,  \label{eq-fobs1}
\end{eqnarray}
where $\Gamma_{\rm Gal} (t_{\rm E}, \theta_{\rm E}, {\bm \pi_{\rm E}})$ is the event rate of a microlensing event 
with parameters $(t_{\rm E}, \theta_{\rm E}, {\bm \pi_{\rm E}})$, 
and $\eta (t_0, u_0, F_{\rm S})$ is proportional to the probability distribution of these three parameters that are 
independent of $t_{\rm E}$, $\theta_{\rm E}$ and ${\bm \pi_{\rm E}}$. The detection efficiency is given by
$\epsilon (t_{\rm E}, t_0, u_0, F_{\rm S})$, and it is effectively independent of 
$\theta_{\rm E}$ and ${\bm \pi_{\rm E}}$ for the events in the {\it Spitzer} sample. Note that 
in other contexts \citep[e.g.,][]{suzuki16} it is common to average over the dependence of $\epsilon$ on 
$t_0$, $u_0$, and $F_{\rm S}$, but for this analysis we consider a specific sample of events for which these
parameters have been measured.

If we want to compare an observed Einstein radius crossing time ($t_{\rm E}$) distribution with the 
predictions from Galactic models, we need to calculate the average detection efficiency as 
a function of $t_{\rm E}$ by simulating event detection processes using artificial events 
\citep{macho_lmc_yr1,macho_lmc_yr2,macho_blg_dia_tau,macho_lmc5.7,sumi03,sum11,mro17}.
However, our interest here is not in the $t_{\rm E}$ distribution, but in the $\pi_{\rm E}$ distribution 
which is obtained by the {\it Spitzer} microlensing parallax measurements.
In this case, we can compare the observed $\pi_{\rm E}$ distribution with the model-predicted distribution 
without any calculation of the detection efficiencies. 
When we consider a specific event with observed parameters 
$(t_{\rm E}, t_0, u_0, F_{\rm S}) = (t_{\rm E, obs}, t_{\rm 0, obs}, u_{\rm 0, obs}, F_{\rm S, obs})$, the 
probability distribution for $\theta_{\rm E}$ and ${\bm \pi_{\rm E}}$ is given by
\begin{eqnarray}
f_{\rm obs} (\theta_{\rm E}, {\bm \pi_{\rm E}} \, | \,  t_{\rm E, obs}, t_{\rm 0, obs}, u_{\rm 0, obs}, F_{\rm S, obs}) & = & \frac{f_{\rm obs} (t_{\rm E, obs}, \theta_{\rm E}, {\bm \pi_{\rm E}}, t_{\rm 0, obs}, u_{\rm 0, obs}, F_{\rm S, obs})}{f_{\rm obs} (t_{\rm E, obs}, t_{\rm 0, obs}, u_{\rm 0, obs}, F_{\rm S, obs})} \nonumber\\
& \propto & \Gamma_{\rm Gal} (\theta_{\rm E}, {\bm \pi_{\rm E}} \, | \,  t_{\rm E, obs})\ , \label{eq-fobs3}
\end{eqnarray}
since we are considering the case of fixed $t_{\rm E}$, $t_0$, $u_0$, and $F_{\rm S}$ at the observed values. 
We have defined the probability density distribution $f (A \, | \,  B)$ to be the conditional probability density 
distribution for $A$ given $B$, and generally $f (A \, | \,  B) = f(A, B)/ f(B)$. In other words, 
$\Gamma_{\rm Gal} (\theta_{\rm E}, {\bm \pi_{\rm E}} \, | \,  t_{\rm E})$ is the probability distribution 
of $\theta_{\rm E}$ and ${\bm \pi_{\rm E}}$ for events with a given $t_{\rm E}$ value. 
We calculate this probability using the Galactic model. 
In Eq. (\ref{eq-fobs3}), the detection efficiency factor is canceled because the values 
of $\theta_{\rm E}$ and ${\bm \pi_{\rm E}}$ are completely
independent of $t_{\rm 0, obs}$ and $u_{\rm 0, obs}$. 
We show below in Section \ref{sec-gammaDs} that our
results do not depend on the weak dependence of $\theta_{\rm E}$ and ${\bm \pi_{\rm E}}$ on $F_{\rm S, obs}$.
The remaining parameter
that depends on $\epsilon (t_{\rm E}, t_0, u_0, F_{\rm S})$ is $t_{\rm E}$, but this 
is fixed to be $t_{\rm E, obs}$ in Eq. (\ref{eq-fobs3}), so we don't need to use the detection efficiency here.
This equation indicates that the observed distribution can be directly compared with the Galactic model.

Because the angular Einstein radius $\theta_{\rm E}$ is not measured for most of the {\it Spitzer} events, 
we focus on the magnitude of microlens parallax $\pi_{\rm E}$ measured by
 the {\it Spitzer} campaign.
In this case, the equation 
\begin{eqnarray}
f_{\rm obs} (\pi_{\rm E} \, | \,  t_{\rm E, obs}, t_{\rm 0, obs}, u_{\rm 0, obs}, F_{\rm S, obs}) \propto  \Gamma_{\rm Gal} (\pi_{\rm E} \, | \,  t_{\rm E, obs}) \label{eq-fobs4}
\end{eqnarray}
is still  true because of the same logic. 
We use the Galactic models explained in Section \ref{sec-model} as the right-hand side of the model-predicted $\pi_{\rm E}$ distribution.
For the left-hand side of observational $\pi_{\rm E}$ distribution, we use the posterior $\pi_{\rm E}$ distribution obtained by combining the Galactic prior with
the raw sample 50 events of \citet{zhu17}, which are described in Section \ref{sec-sample}.
Note that our choice of focusing on the 1D distribution of $\pi_{\rm E}$ rather than the 2D distribution of $\pi_{\rm E,N}$ and $\pi_{\rm E,E}$ 
is because there is no well-established statistical test to compare 2D distributions as long as we know.
Using only 1D distribution makes our results more conservative.

\section{Models} \label{sec-model}
To calculate $\Gamma_{\rm Gal} (\pi_{\rm E} \, | \,  t_{\rm E, obs})$, we need a Galactic model, which consists 
of the stellar mass function, stellar density distribution, and velocity distribution in our galaxy.
Microlensing groups have developed a number of such models, and they are often referred to
as ``standard Galactic model" \citep{sum11, ben14, zhu17, mro17, jun18}. We use the model
presented by \citet{zhu17} in the paper that presented this {\it Spitzer} sample, as well as
Galactic models presented by \citet{sum11} and \citet{ben14}
for our comparison with this {\it Spitzer} microlensing parallax sample.
Hereafter we refer to these papers as and models as Z17, S11 and B14, respectively. 

The Z17 and S11 models are based on the Galactic model developed by \citet{han95}, while the B14 model 
is based on the Galactic model developed by \citet{rob03} and it includes a central hole in the disk that
was created by the disk instability thought to have formed the central Galactic bar, as well as bar rotation.
The B14 model also includes a thick disk and spheroid, but none of these features are considered in
\citet{han95}.
In this section we give the outline of these three models focusing on the differences between them.
We summarize them in Table \ref{tab-models}.
More details are found in each paper and references therein.

\begin{deluxetable}{lllll}
\tablecaption{Summary of the three Galactic models used in this work. \label{tab-models}}
\tablehead{
\colhead{} & \colhead{Model} & \colhead{Z17} & \colhead{S11} & \colhead{B14}
}
\startdata
Stellar          & Initial Mass function                                            & Eq. (\ref{eq-MF})                &  Eq. (\ref{eq-MF})                      & Eq. (\ref{eq-MF})              \\      
population   &  \ \ \ \  $(\alpha_{\rm hm}, \alpha_{\rm ms}, \alpha_{\rm bd})$        & (2.3, 1.3, 0.3)                  &  (2.0, 1.3, 0.5)                        & (2.0, 1.3, 0.5)                \\
                   &  \ \ \ \  $(M_{\rm max}, M_1, M_2, M_{\rm min})~[M_{\odot}]$           & $(8.0, 0.50, 0.08, 10^{-5})$     &  $(8.0, 0.70, 0.08, 10^{-5})$           & $(8.0, 0.70, 0.08, 10^{-5})$   \\
                   & Age and Metalicity                                               & B18\tablenotemark{a}                    &  B18\tablenotemark{a}      & B18\tablenotemark{a}    \\      
                   & $\gamma$ of event rate $\propto D_S^{-\gamma}$                   & 2.85                             &   2                                     & 1.5                            \\      
Bulge          & Density                                                          & Eq. (\ref{eq-rhobS11})           &  Eq. (\ref{eq-rhobS11})                 & Eq. (\ref{eq-rhobB14})         \\      
structure      &  \ \ \ \ Bar angle $\alpha_{\rm bar}$ [deg.]                           & $30$                             & $20$                                    & $20$                           \\
                   &  \ \ \ \ $\rho_{\rm B, 0}$ $[M_{\odot}\,{\rm pc}^{-3}]$                & 3.76\tablenotemark{b}            & 2.07                                    & 2.07                           \\
                   &  \ \ \ \ $(x_0, y_0, z_0)$ [pc]                                        & (1590, 424, 424)                 & (1580, 620, 430)                        & (1580, 620, 430)               \\
                   & Mean velocity                                                    &  0~km/s                          & $50~{\rm km/s}$ (stream)             & $50~{\rm km/s/kpc}$ (rot.) \\
                   &  \ \ \ \  Dispersion [km/s]                                            & (120, 120, 120)\tablenotemark{c} & (113.6, 77.4, 66.3)\tablenotemark{c}    & (114.0, 103.8, 96.4)\tablenotemark{d}\\
Disk            & Density                                                          & Eq. (\ref{eq-rhodS11})           &  Eq. (\ref{eq-rhodS11})                 & Eq. (\ref{eq-rhodB14})         \\      
structure     &  \ \ \ \ Local density\tablenotemark{e} $[M_{\odot}\,{\rm pc}^{-3}]$   & 0.038\tablenotemark{b}           & 0.06                                    & 0.039                          \\
                   &  \ \ \ \ Disk scale length, height [pc]                                & (3500, 325)                      & (3500, 325)                             & (2530, 200)                    \\
                   &  \ \ \ \ Hole scale length, height [pc]                                &  No hole                         &  No hole                                & (1320, 104)                    \\
                   & Rotation speed                                                   &  240~km/s                        &  220~km/s                               & 218.0~km/s                     \\      
                   &  \ \ \ \  Dispersion [km/s]                                            & (33, 18)\tablenotemark{f}        & (30, 30)\tablenotemark{f}               & (27.9, 19.1)\tablenotemark{f}  \\
Sun                & Location $(R_{\odot}, z_{\odot})$ [pc]                           & (8300,  27)                      & (8000,   0)                             & (8200,   0)                    \\      
                   & Velocity $(v_{\odot,y}, v_{\odot,z})$ [km/s]                     & (252, 7)                         & (220, 0)                                & (242, 7.25)                    \\\hline 
\enddata
\tablenotetext{a}{\citet{ben18}}
\tablenotetext{b}{Converted from the original values of number density $(n_{\rm B,0}, n_{\rm D,0}) = (13.7, 0.14)~{\rm pc}^{-3}$. See section \ref{sec-den} for detail.}
\tablenotetext{c}{Velocity dispersion along $(x', y', z')$ axis.}
\tablenotetext{d}{Velocity dispersion along $(x, y, z)$ axis.}
\tablenotetext{e}{Stellar volume density around the Sun location, which is equivalent to $\rho_{\rm D, 0}$ for Z17 and S11 models.}
\tablenotetext{f}{Velocity dispersion along $(y, z)$ axis.}
\tablecomments{Small modifications, such as $M_{\rm max}$ and $M_{\rm min}$ values, are adopted in each model compared to the original ones.}
\end{deluxetable}

\subsection{Mass function}
All the three models use a broken power-law form for the stellar mass function for main sequence stars, 
and the stellar mass functions are assumed to be continuous at the breaks. However, it is also
important to consider the possibility of microlensing by brown dwarfs and stellar remnants. 
The possibility that the lens may be a stellar remnant is often ignored for planetary events, because stellar
remnants are thought to rarely host planets, but we cannot neglect this possibility for this analysis 
because the \citet{zhu17} sample consists of single lens events.
Also, the star formation process does not distinguish between low-mass stars and brown dwarfs, so 
we consider mass functions whose slope on brown dwarf mass region extended down to planetary masses.
However, the low-mass tail of the mass function has little influence on our results as the 
\citet{zhu17} sample is biased toward longer $t_{\rm E}$ events.

We consider the present-day mass function as follows.
First we take the initial mass function (IMF) to be
\begin{equation}
\frac{dN}{dM} \propto
\begin{cases}
M^{-\alpha_{\rm hm}}  & \text{ when $M_1 < M < M_{\rm max}$} \\
M^{-\alpha_{\rm ms}}  & \text{ when $M_2 < M < M_1$} \\
M^{-\alpha_{\rm bd}}  & \text{ when $M_{\rm min} < M < M_2$} \ .  \label{eq-MF}
\end{cases}
\end{equation}
Z17 uses the values $(M_1, M_2) = (0.50, 0.08)\, M_{\odot}$ and 
$(\alpha_{\rm hm}, \alpha_{\rm ms}, \alpha_{\rm bd}) = (2.3, 1.3, 0.3)$ following \citet{kro01}, while S11 uses 
$(M_1, M_2) = (0.70, 0.08)\, M_{\odot}$ and $(\alpha_{\rm hm}, \alpha_{\rm ms}, \alpha_{\rm bd}) = (2.0, 1.3, 0.5)$ 
based on a comparison with the observed $t_{\rm E}$ distribution from their microlensing survey. 
B14 also uses the S11 mass function. We use a minimum mass of 
$M_{\rm min} = 10^{-5} M_{\odot}$ for all the three models, but this has little effect because planetary masses
are strongly disfavored by the large $t_{\rm E}$ values of the Z17 sample.
We adopt $M_{\rm max} = 8.0~M_{\odot}$ as the maximum mass of the IMF and ignore stars with 
initial masses of $> 8.0~M_{\odot}$ that will have evolved into neutron stars and black holes.
We  construct the present-day mass function by randomly selecting a star from our IMF, given in 
equation~\ref{eq-MF}, and then randomly selecting an age and metalicity from the relatively wide
distribution used by \citet{ben18}. Stellar magnitudes are determined with the 
PARSEC isochrones \citep{bre12, che14, tan14}, and for stars that have evolved into white dwarfs, we use
the  initial-final mass relation of \citet{elb18} to determine the final white dwarf masses.
\citet{zhu17} also considered another mass function of the form $dN/dM \propto M^{-1}$, but we
 do not use this model.

\subsection{Density distribution} \label{sec-den}
The Z17 and S11 models use the boxy-shaped bulge model of \citet{dwe95},
\begin{eqnarray}
\rho_{\rm B} =  \rho_{\rm B, 0} \exp (-0.5\ r_s^2); \hspace{0.3cm} r_s =  
\left \{ \left [ \left ( \frac{x'}{x_0} \right)^2 + \left( \frac{y'}{y_0} \right)^2 \right ]^2 + \left( \frac{z'}{z_0} \right)^4 \right \}^{1/4},  \label{eq-rhobS11}
\end{eqnarray}
and the double exponential disk model of \citet{bah86},
\begin{eqnarray}
\rho_{\rm D} =  \rho_{\rm D, 0} \exp \left[- \left( \frac{R-R_\odot}{h_R} + \frac{|z|}{h_z} \right) \right] ; 
\hspace{0.3cm} R = \sqrt{x^2 + y^2},  \label{eq-rhodS11}
\end{eqnarray}
and they use $\rho = \rho_{\rm B} + \rho_{\rm D}$ as the total density distribution, without including a separate thick disk
or spheroid component. We use $(x, y, z)$ to refer to galactocentric coordinate and  $(x', y', z')$ to refer to
a coordinate system that is rotated about the $z$-axis aligned by an angle $\alpha_{\rm bar}$ so that
the $x'$ axis is aligned with the Galactic bar, The Z17 model uses the following parameters:
$(\rho_{\rm B, 0},  \rho_{\rm D, 0}) = (3.76, 0.038) \, M_{\odot} \, {\rm pc}^{-3}$,  
$(x_0, y_0, z_0) = (1590, 424, 424) \, {\rm pc}$, $\alpha_{\rm bar} = 30^{\circ}$ and $R_\odot = 8300 \, {\rm pc}$.
The S11 model uses somewhat different parameters: 
$(\rho_{\rm B, 0},  \rho_{\rm D, 0}) = (2.07, 0.06)\, M_{\odot} \, {\rm pc}^{-3}$, 
$(x_0, y_0, z_0) = (1580, 620, 430) \, {\rm pc}$, $\alpha_{\rm bar} = 20^{\circ}$ and $R_\odot = 8000 \, {\rm pc}$.
Both the Z17 and S11 models use the same disk scale length and scale height, $(h_R, h_z) = (3500, 325)\, {\rm pc}$.
The mass density values $(\rho_{\rm B, 0},  \rho_{\rm D, 0})$ for the Z17 models were derived from the original 
number density values of $(n_{\rm B, 0}, n_{\rm D, 0}) = (13.7, 0.14) ~  {\rm pc}^{-3}$, but the original 
number density values are used for our calculations.

The B14 model employs a modified boxy-shaped bulge model of \citet{rob03} with a density given by
\begin{equation}
\rho_{\rm B} =  
\begin{cases}
\rho_{\rm B, 0} \exp (-0.5\, r_s^2) & \text{ when $R < 2400$ pc} \\
\rho_{\rm B, 0} \exp (-0.5\, r_s^2) \times \exp \left[ -0.5\, \left( \frac{R - {\rm 2400~pc}}{{\rm 500 pc}}\right)^2 \right]  & \text{ when $R > 2400$ pc} \ .  \label{eq-rhobB14}
\end{cases}
\end{equation}
The B14 disk model has a central hole that is expected due to the formation of the bar-shaped bulge 
from disk instability. This minimizes, but does not completely remove, an unphysical feature of the S11 and 
Z17 models, which have a  singular velocity field at Galactic longitude $l = 0$ at the distance of the Galactic center. 
This can lead to unrealistic conclusions for lines of sight close to $l = 0$.

The B14 disk model  is given by
\begin{equation}
\rho_{\rm D} = \rho_{\rm D, 0} \left\{ \exp \left[-\sqrt{0.5^2 + \frac{a^2}{h^2_{R_+}}} \, \right]  - \exp \left[-\sqrt{0.5^2 + \frac{a^2}{h^2_{R_-}}} \, \right] \right \} ; \hspace{0.3cm} a^2 = R^2 + \left( \frac{z}{0.079} \right)^2 \ . \label{eq-rhodB14}
\end{equation}
The B14 model uses $(\rho_{\rm B, 0},  \rho_{\rm D, 0}) = (2.07, 1.10)\, M_{\odot} \, {\rm pc}^{-3}$, 
$(x_0, y_0, z_0) = (1580, 620, 430) \, {\rm pc}$, $\alpha_{\rm bar} = 20^{\circ}$ and 
$(h_{R_+}, h_{R_-}) = (2530, 1320)~{\rm pc}$.
Also they use $R_\odot = 8200 \, {\rm pc}$ as the distance to the Sun from the Galactic center.
The B14 model also includes two Galactic components that are ignored by the other models. These are
the thick disk, with density  $\rho_{\rm td}$, and the spheroid with density, $\rho_{\rm sph}$, following \citet{rob03}.
The total density in the B14 model is then given by $\rho = \rho_{\rm B} + \rho_{\rm D} + \rho_{\rm td} + \rho_{\rm sph}$.
Note that contributions from $\rho_{\rm td}$ and especially $\rho_{\rm sph}$, are usually quite small, but they
can make an important contribution to events with high relative lens-source proper motions. In fact, there is at
least one well measured microlensing event confirmed to be due to a thick-disk lens star \citep{gould09}.

Although S11 and B14 use the same $\rho_{\rm B, 0}$ and $(x_0, y_0, z_0)$ values, the total bar mass 
is $1.8 \times 10^{10}~M_{\odot}$ for S11 model while it is $1.65 \times 10^{10}~M_{\odot}$ for B14 model because 
the bar density model of B14 has an additional term reducing the density at $R > 2400~{\rm pc}$.
Also note that the $\rho_{\rm D, 0}$ value for B14 model is a value near the Galactic center without 
the hole, in contrast to the $\rho_{\rm D, 0}$ values for S11 and Z17 models at the Sun location.
The B14 disk model gives $0.039~M_{\odot} \, {\rm pc}^{-3}$ as the density value at the Sun location.

\subsection{Velocity Distribution}
The velocity distribution is characterized by the observer's 
transverse velocity and the mean transverse velocities and
dispersions for all components of the Galaxy.
The Sun's velocity and the velocity distribution for the disk stars are similar with each other among the three 
models we consider, as summarized in Table \ref{tab-models}. 
For the mean velocity of bulge stars, while Z17 applies $0\,{\rm km/s}$ for all directions, S11 applies a streaming 
velocity with $50\, {\rm km/s}$ along $x'$ axis and 
B14 applies a rigid body rotation of the bar with the angular velocity of $50 \, {\rm km/s/kpc} $.
For the velocity dispersion of bulge stars, Z17 uses 
$(\sigma_{v_x'}, \sigma_{v_y'}, \sigma_{v_z'}) = (120, 120, 120)\, {\rm km/s}$ for the velocity dispersion 
along $x'$, $y'$ and $z'$ axes and 
S11 uses $(\sigma_{v_x'}, \sigma_{v_y'}, \sigma_{v_z'}) = (113.6, 77.4, 66.3)\, {\rm km/s}$.
Also B14 uses $(\sigma_{v_x}, \sigma_{v_y}, \sigma_{v_z}) = (114.0, 103.8, 96.4)\, {\rm km/s}$ for the 
velocity dispersion along $x$, $y$ and $z$ axes.

\subsection{Event rate}\label{sec-everate}
The microlens event rate, $\Gamma_{\rm Gal} (t_{\rm E}, \theta_{\rm E}, {\bm \pi_{\rm E}})$, can be calculated 
numerically by picking a combination of a source star and a lens star both following the Galactic model distribution,
as discussed above. This must be weighted by a factor,  $2 D_L^2 \theta_{\rm E} \mu_{\rm rel} D_S^{2-\gamma}$,
that is proportional to the event rate. The factors $2 D_L^2 \theta_{\rm E} \mu_{\rm rel}$, $D_S^2$, and 
$D_S^{-\gamma}$ account for the area swept per unit time by the Einstein ring of the selected lens, the 
increase in volume with increasing distance, and decreasing number of source stars which have detectable 
brightness with increasing distance, respectively\citep{kir94}. This $D_S^{-\gamma}$ factor is a rather crude 
approximation of the actual distance dependence of source stars, since the real dependence is a complicated
function of source magnitude and position on the sky. \citet{ben18, bennett_mb10117} presented a much
more accurate method, but this becomes quite complicated for large samples of events. The models
we consider use different $\gamma$ values. The B14, S11, and Z17 models use  $\gamma = 1.5$, $\gamma = 2$
and $\gamma = 2.85$, respectively.

\begin{figure}
\begin{minipage}{0.32\hsize}
\centering
\includegraphics[width=62mm]{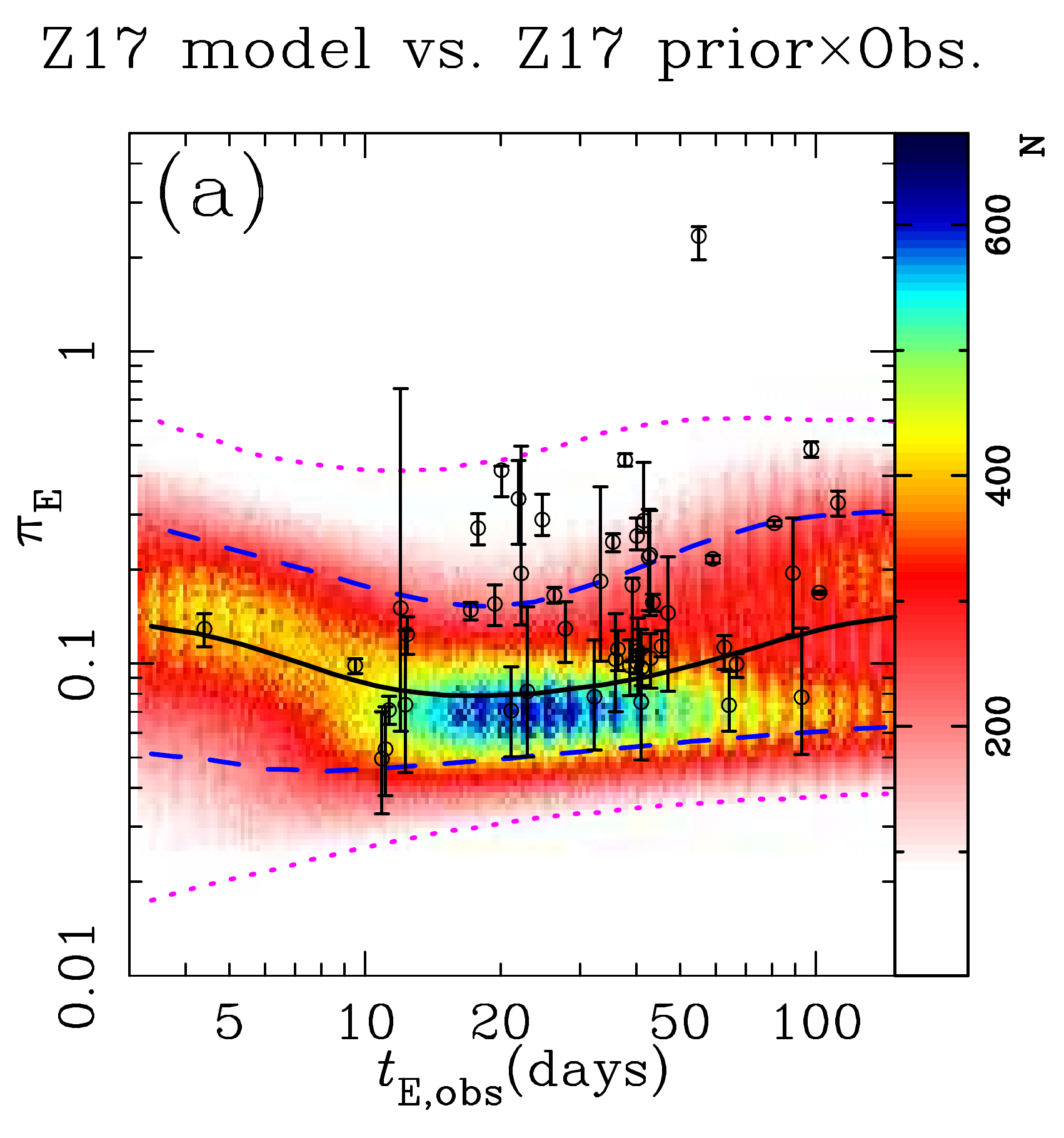}
\end{minipage}
\begin{minipage}{0.32\hsize}
\centering
\includegraphics[width=62mm]{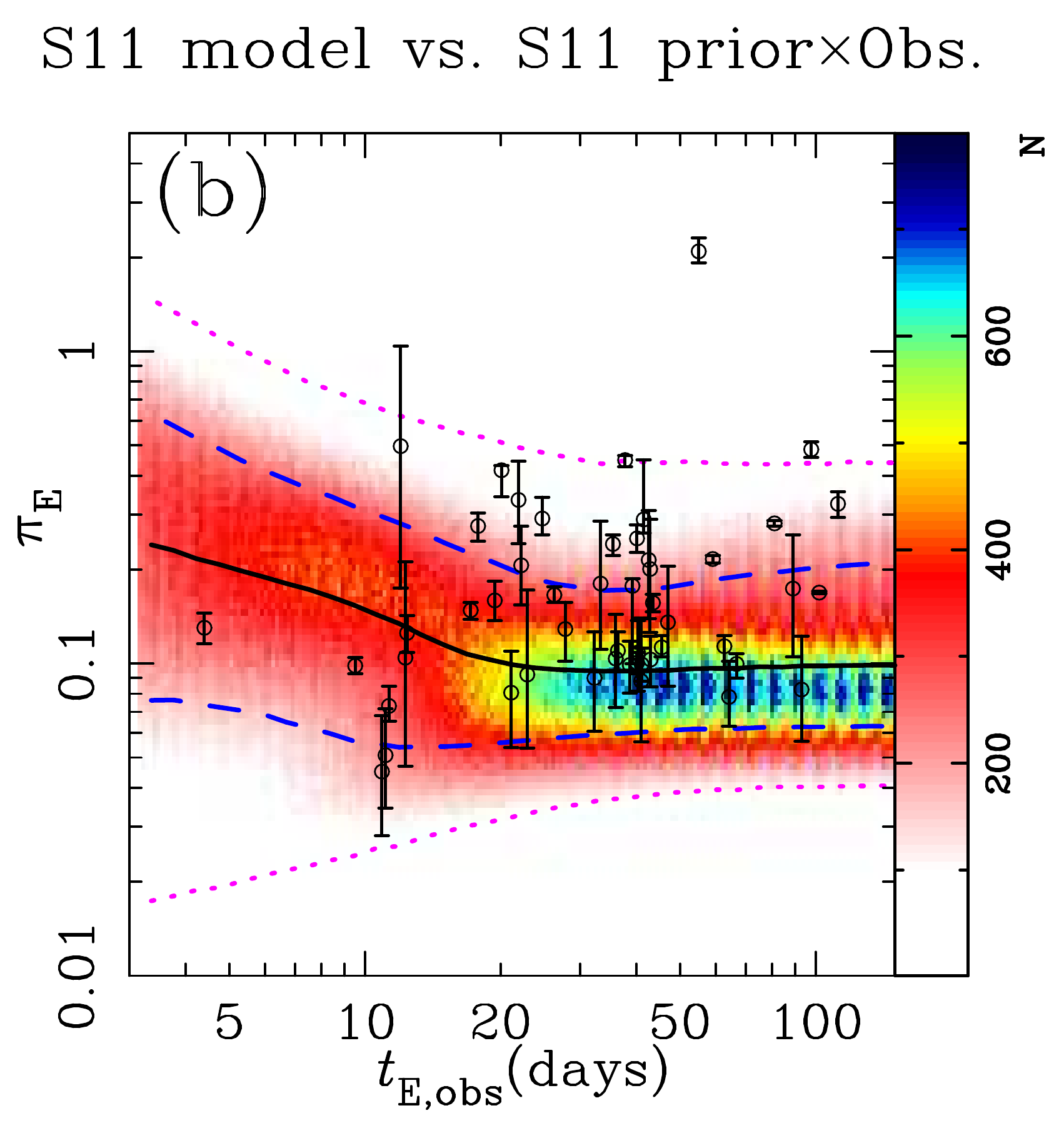}
\end{minipage}
\begin{minipage}{0.32\hsize}
\centering
\includegraphics[width=62mm]{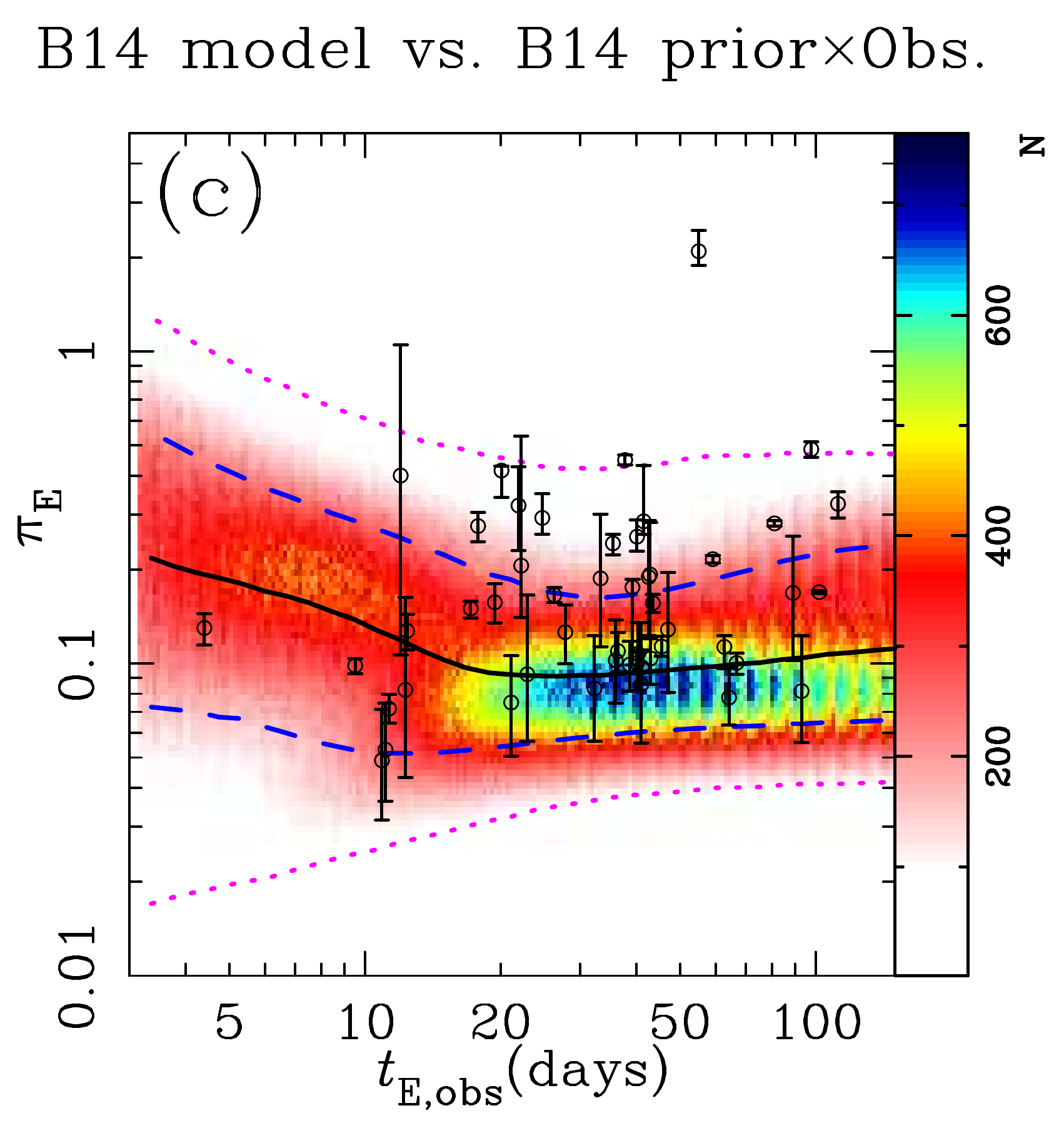}
\end{minipage}

\caption{Comparison of the observations with the Galactic model. Each panel shows a different Galactic model.
Open circles are the median, $\widetilde{ \pi}_{\rm E, post}$, and  68\% confidence intervals of the 
posterior distribution $f_{\rm post} (\pi_{\rm E})$ for the 50 measurements from 2015 {\it Spitzer} microlensing 
campaign \citep[the ``raw" sample of][]{zhu17} convolved with the prior from each Galactic model.
The color maps indicate the event rate, $\Gamma_{\rm Gal} (\pi_{\rm E} \, | \,  t_{\rm E, obs})$. 
$10^5$ artificial events are generated for each bin of $t_{\rm E,obs}$ with the width 0.05 dex. The
black solid, blue dashed, and magenta dotted lines indicate the median, $1~\sigma$, and $2~\sigma$
for $\Gamma_{\rm Gal} (\pi_{\rm E} \, | \,  t_{\rm E, obs})$, respectively.}
\label{fig-tEvtil}
\end{figure}

The calculated event rate, $\Gamma_{\rm Gal} (\pi_{\rm E} \, | \,  t_{\rm E, obs})$, is shown as a function of $ t_{\rm E, obs}$
as the color maps in Figure~\ref{fig-tEvtil} for the Z17, S11 and B14 models. For this plot, the event rates were
calculated over the range $0.50 < \log (t_{\rm E, obs}/{\rm days}) < 2.20$, by dividing this range into 34 bins
of width 0.05 dex and then generating $10^5$ artificial events, with simulated $\pi_{\rm E}$ values, in each bin 
for each of our 3 models. We select a typical Galactic coordinate of $(l, b) = (1.0^{\circ}, -2.2^{\circ})$ for this
sample to use for these calculations for this plot. Note that the coordinate of each event is used in calculations for statistical tests in Section \ref{sec-test}.

\section{Microlensing Event Sample} \label{sec-sample}
We use the 50 single microlens events discovered by OGLE-IV survey and observed by 
the 2015 {\it Spitzer} campaign \citep{zhu17}, as our event sample. This is the raw sample of \citet{zhu17}.
The $\pi_{\rm E}$ distribution of this sample follows
$f_{\rm obs} (\pi_{\rm E} \, | \,  t_{\rm E, obs}, t_{\rm 0, obs}, u_{\rm 0, obs}, F_{\rm S, obs})$
and satisfies Eq. (\ref{eq-fobs4}) when the following two assumptions are true:
\begin{enumerate}
\item The measured $t_{\rm E}$ and $\pi_{\rm E}$ are both randomly distributed around the true values 
of those parameters.
\item The event selection process produces no bias in the $\pi_{\rm E}$ distribution of the sample.
\end{enumerate}

Assumption 2 is the reason why we use the \citet{zhu17} raw sample instead of their final sample.
The \citet{zhu17} final sample includes only events where $\pi_{\rm E}$ can be measured, which they define as
events with $\sigma (D_{8.3}) \leq 1.4~$kpc, where $D_{8.3} = {\rm kpc}/(\pi_{\rm rel}/{\rm mas} + 1/8.3)$.
This selection clearly violates assumption 2, since events with small $\pi_{\rm E}$ are much more likely to
have large $\sigma (D_{8.3})$ values. In contrast, their raw sample was 
selected independently of the measured $\pi_{\rm E}$ values, so this selection should not introduce
bias into the $\pi_{\rm E}$ distribution of the raw sample.
Note that the  {\it Spitzer} target selection procedure \citep{yee15} does favor bright sources, which
could violate assumption 2. We show that this does not affect our conclusions 
in Section \ref{sec-gammaDs}.

\subsection{Applying the Galactic Prior to Parallax Measurements} \label{sec-galpri}

Microlensing parallax measurements typically have large uncertainties because the effect is difficult to 
measure. Microlensing parallax measurements by satellites, like {\it Spitzer}, typically 
have multiple degenerate solution \citep{ref66, gou94}, and microlensing parallax signals due to the orbital motion of the
Earth generally have very large uncertainties in the direction perpendicular to the acceleration of Earth
at the time of the event \citep[e.g.,][]{mur11, bha18}. Even when assumption 1 is true, these large uncertainties mean that 
the assumed prior distribution can have an important influence on the inferred $\piEbold$ distribution.
In particular, if no prior is applied, it is equivalent to applying a uniform prior, and this can lead to
overestimates of $\pi_{\rm E}$ in situations where the true $\pi_{\rm E}$ values are near the limits
of the measurement method. 
\citet{zhu17} apply the Galactic prior to derive the distributions of the lens mass and 
distance, but the reported $\piEbold$ values are the one with the uniform prior.

We apply the Galactic prior to generate the following posterior distribution for each event
\begin{eqnarray}
f_{\rm post} (\pi_{\rm E,N}, \pi_{\rm E,E}) \propto \sum_{i} \Gamma_{\rm Gal} (\pi_{\rm E, N}, \pi_{\rm E, E} \,  | \,  t_{{\rm E, obs}, i}) & \nonumber\\ 
\times {\cal N} (\pi_{\rm E, N}; \pi_{{\rm E, N, obs}, i}, \sigma^2_{\pi_{\rm E, N}, i}) &  {\cal N} (\pi_{\rm E, E}; \pi_{{\rm E, E, obs}, i} , \sigma^2_{\pi_{\rm E, E}, i}) e^{-\Delta\chi^2_i/2} ,  \label{eq-fpost2}
\end{eqnarray}
where the summation is conducted over the degenerate solutions for each event. 
${\cal N} (x; \, {\overline x}, \sigma^2_{x})$ is the Gaussian distribution with the mean of ${\overline x}$
and the standard deviation of the measurement error $\sigma_{x}$, where we use the best-fit 
values and error-bars for $\pi_{\rm E, N}$ and $\pi_{\rm E, E}$ reported by \citet{zhu17} as the mean and error, respectively. 
The Einstein radius crossing
times for each of the degenerate solutions, $t_{{\rm E, obs}, i}$, are nearly identical, so 
we calculate the posterior distribution with one representative $t_{\rm E, obs}$ value for each event.
Note that \citet{zhu17} apply both the Galactic prior and the ``Rich" argument \citep{cal15}, which gives a prior of $\pi_{\rm E}^{-2}$, to their derivation of
the distribution of lens masses and distances. This is incorrect as the ``Rich" argument is just a crude
attempt to apply a prior to the $\pi_{\rm E}$ measurement and it is already included
in the Galactic prior. A more recent paper \citet{rich_arg}, inspired by an early version of this paper,
confirms our identification of the ``Rich" argument as a Galactic prior.

\begin{figure}
\centering
\includegraphics[width=6.3in]{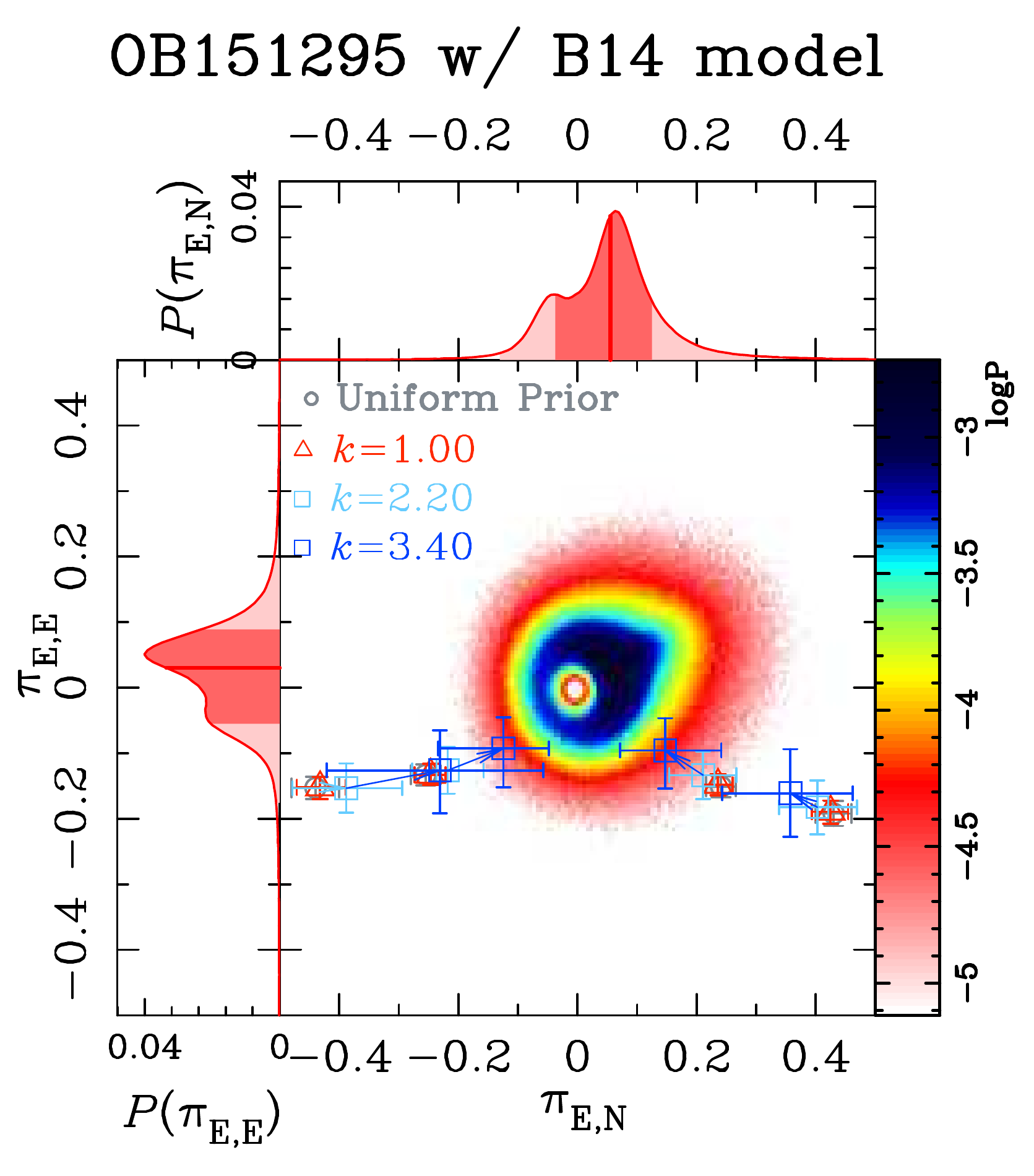}
\caption{The color coding indicates the two dimensional microlensing parallax prior distribution for event
OGLE-2015-BLG-1295 according to the B14 model. The prominent extension in the NNE
direction where the probability is highest is the direction of Galactic disk rotation. This is the
preferred microlensing parallax direction for lens stars in the Galactic disk. The red boxes with error bars
are the median and 68\% confidence interval for the posterior distributions for the 4 degenerate
{\it Spitzer} microlensing parallax measurements for this event. The cyan and blue boxes with error bars 
indicate the median and  68\% confidence intervals for the posterior distribution for the
{\it Spitzer} microlensing parallax values
when the error bars are inflated by factors of $k = 2.2$ and 3.4, respectively. The measured values
seem consistent with the prior only for the largest error bar inflation factor of $k = 3.4$.}
\label{fig-piE2d}
\end{figure}

An example of how the application of this prior changes the inferred posterior distribution is shown 
in Figure~\ref{fig-piE2d}. This figure shows the  $\piEbold$ prior distribution for event OGLE-2015-BLG-1295, an event
with $t_{\rm E} \simeq 42\,$days, and this figure indicates that microlensing parallax values in the range of 
$0.05 \simlt \pi_{\rm E} \simlt 0.2$ are favored. There is also a strong enhancement of the prior probability
in the NNE direction, particularly at larger $\pi_{\rm E}$ values. This is the direction of Galactic disk rotation,
which is preferred because of the large number microlensing events due to bulge source stars
and lens stars orbiting in the disk.  Also, note that values of $\pi_{\rm E}< 0.01$ 
are strongly disfavored, contrary to what the ``Rich" argument would imply. 

Figure~\ref{fig-piE2d} also shows the measured $\piEboldobs$ values as grey circles, as well as the 
centroid of the posterior probability distribution obtained by convolving this prior with a 2-dimensional Gaussian 
describing the measurement and its error bars. The medians of the posterior distributions for 
the 4 degenerate solutions are indicated
by red triangles with error bars indicating the 68\% confidence interval in both directions. (Note that the red 
triangles are located in almost exactly the same place as the grey circles for this event.)

We now apply the prior to all 50 of the events in the \citet{zhu17} sample, and then, we
convert the two dimensional posterior distribution for $\piEbold$, $f_{\rm post} (\pi_{\rm E,N}, \pi_{\rm E,E})$,
into the posterior distribution for the length of the parallax vector, $\pi_{\rm E}$. This posterior distribution is therefore
\begin{eqnarray}
f_{\rm post} (\pi_{\rm E})  = \int_{\phi_{\rm E}} f_{\rm post} (\pi_{\rm E}, \phi_{\rm E})\, d\phi_{\rm E}\label{eq-fpost1}
\end{eqnarray}
where $\phi_{\rm E}$ is the direction angle of ${\bm \pi_{\rm E}}$ vector. and then use 
for the statistical test.
The black open circles in each panel of Figure~\ref{fig-tEvtil} show the median, $\widetilde{ \pi}_{\rm E, post}$, and 68\% confidence intervals of $f_{\rm post} (\pi_{\rm E})$ 
as a function of the observed values of $t_{\rm E, obs}$ for all 50 events in the Z17 sample.
For the prior distribution in each of $f_{\rm post} (\pi_{\rm E})$ calculation, we use the compared Galactic model in each panel with 
parameters for each event (coordinate, Earth projected velocity at $t_{\rm 0, obs}$, etc.). 
The difference between the different Galactic priors are responsible for the different distributions the black open circles 
in different panels of Figure~\ref{fig-tEvtil}.

At first glance, only one event is an obvious outlier compared to the 
model distribution in Figure~\ref{fig-tEvtil}\footnote{This outlier is OGLE-2015-BLG-1227, where the {\it Spitzer} data seems to only 
cover the baseline.}, and most of the other events' $\widetilde{ \pi}_{\rm E, post}$ values are 
within $\pm 2~\sigma$ of the simulated event rate distribution, $\Gamma_{\rm Gal} (\pi_{\rm E} \, | \,  t_{\rm E, obs})$.
As a result, it would be difficult to argue that the $\pi_{\rm E}$ values measured by {\it Spitzer} are too large
on an event-by-event basis. However, when we consider all 50 measurements, we see that most events
(37 for the S11, B14 models and 40 for the Z17 model) have measurement posteriors,
$\widetilde{ \pi}_{\rm E, post}$, above the median values (for each individual $t_{\rm E, obs}$ value). The
probabilities of these outcomes is given by the binomial distribution, and the probability for at least 37 events
above the median is $4.6\times 10^{-4}$, while the probability for $\geq 40$ events above the median
is $1.2\times 10^{-5}$. This 
indicates a strong inconsistency between the $\pi_{\rm E}$ measurements with the Galactic models.
It is also obvious that the observations do not match the models by a visual comparison.

\subsubsection{Error Bar Inflation Factors} \label{sec-inflate}
In Section~\ref{sec-dis}, we will conclude that the discrepancy is mainly due to systematic errors in the {\it Spitzer} microlensing parallax measurements.
As we discuss in Section~\ref{sec-adde}, we feel that the optimal way to deal with this
problem with the {\it Spitzer} microlensing parallax measurements is to try to remove or correct systematic
errors in the {\it Spitzer} photometry. This is beyond the scope of this work, but it is an approach that has been tried for
a few events by \citet{gou20} and \citet{dan20}, inspired, at least in part, but an early version of the work presented
here. However, it is not yet clear how successful these approaches will be when applied to a large sample of events.
So, we consider a simpler approach, in which we attempt to mitigate the effect of the systematic errors by increasing
(or inflating) the error bars of $\piEboldobs$ by a factor $k$, that we refer to as the error bar inflation factor. The idea behind this
is to give an estimate of the effect of these systematic errors on the parallax measurement uncertainties and to 
provide a crude way to correct the measurements in cases where systematic photometry errors cannot be corrected.
As we discuss in Section \ref{sec-test}, this procedure has some drawbacks when $k$ is very large, but we feel that these error bar inflation factors
provide a useful way to characterize the apparent systematic errors.

In Figure~\ref{fig-piE2d},
the red triangles appear to be strongly inconsistent with the prior distribution for this event, so
we also consider the same convolutions of the Galactic prior distributions with Gaussians describing the 
$\piEboldobs$ measurements with inflated error bars. The cyan and blue boxes with error bars indicate
the posterior distributions medians and 68\% confidence intervals when the error bars increased by inflation factors of 
$k = 2.2$ and $k = 3.4$, respectively. An inflation factor of $k = 3.4$ seems to be enough to make two or three of
the degenerate solutions consistent with the prior for this event.

We note that the modification of these parallax values to the median of the posterior values will certainly
change the predicted {\it Spitzer} light curves and they do not follow the original data points.
However, this is expected because the introduction of $k > 1$ is motivated by the idea that 
the original {\it Spitzer} light curve photometry suffered from systematic errors.

We also apply the error bar inflation factor of 3.4 to the comparison between the {\it Spitzer} $\pi_{\rm E}$ 
measurements and the B14 model from Figure~\ref{fig-tEvtil}(c). The result is shown in Figure~\ref{fig-B14_NF}(a),
which indicates that this changes the situation quite dramatically. The obvious discrepancy between models and the
{\it Spitzer} microlensing parallax measurements largely disappears when the measurement errors are increased
by a factor of 3.4.

\section{Statistical Tests} \label{sec-test}

We now introduce the inverse percentile, which is defined as $x(y)$ $(0 \leq x \leq 1)$ 
when $y$ is the $100 x$-th percentile of a given probability density function of $y$, to 
more quantitatively evaluate the mismatch between the {\it Spitzer} microlensing parallax 
measurements and the models.
The observed value of an observable quantity that follows a given distribution should have a uniform 
inverse percentile distribution from 0 to 1.
Therefore, if the parallax measurement for each event follows the predicted Galactic prior, 
$\Gamma_{\rm Gal} (\pi_{\rm E} \, | \,  t_{\rm E, obs})$, the position of the measured $\pi_{\rm E}$ values 
with respect to the Galactic prior for each event,
\begin{equation}
P_{\rm Gal}  (\pi_{\rm E, prior} \geq \pi_{\rm E} \, | \,  t_{\rm E, obs}) \equiv \int_{\pi_{\rm E}}^{\infty} \Gamma_{\rm Gal} (\pi_{\rm E}' \, | \,  t_{\rm E, obs}) \, d \pi_{\rm E}' \ , \label{eq-Pgal}
\end{equation}
should follow a uniform distribution as long as a proper Bayesian prior is included in the determination
of the measured $\pi_{\rm E}$ value. 
$P_{\rm Gal}  (\pi_{\rm E, prior} \geq \pi_{\rm E} \, | \,  t_{\rm E, obs})$ ranges from 0 to 1, depending on 
the measured $\pi_{\rm E}$ value. 
A $\pi_{\rm E}$ value that gives $P_{\rm Gal}  (\pi_{\rm E, prior} \geq \pi_{\rm E} \, | \,  t_{\rm E, obs}) = x$
is the $100 x$-th percentile of the Galactic prior $\Gamma_{\rm Gal} (\pi_{\rm E} \, | \,  t_{\rm E, obs})$, so
we refer to $P_{\rm Gal}  (\pi_{\rm E, prior} \geq \pi_{\rm E} \, | \,  t_{\rm E, obs})$ as the inverse percentile in this paper.

We use the Anderson-Darling (AD) test, which is more sensitive than the commonly used 
Kolmogorov-Smirnov (KS) test,
to compare the inverse percentile distribution from the {\it Spitzer} microlensing parallax
measurements to the uniform distribution.
For each inverse percentile calculation, we use the parameters for each event, which include the event's Galactic coordinates and the Earth's velocity at the
time of the event peak, $t_{\rm 0, obs}$.

\begin{figure}
\begin{minipage}{0.45\hsize}
\centering
\includegraphics[width=82mm]{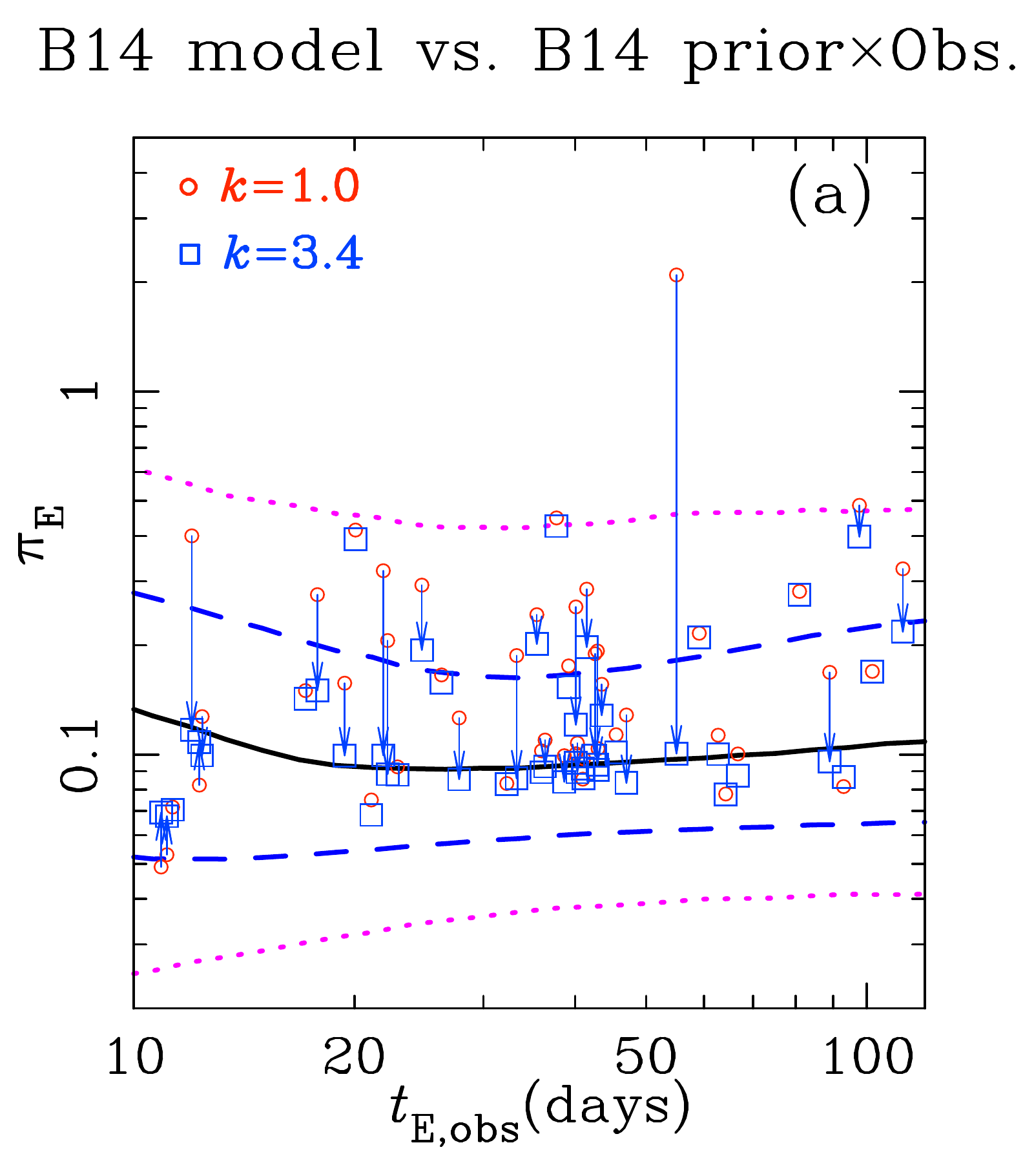}
\end{minipage}
\begin{minipage}{0.55\hsize}
\centering
\includegraphics[width=95mm]{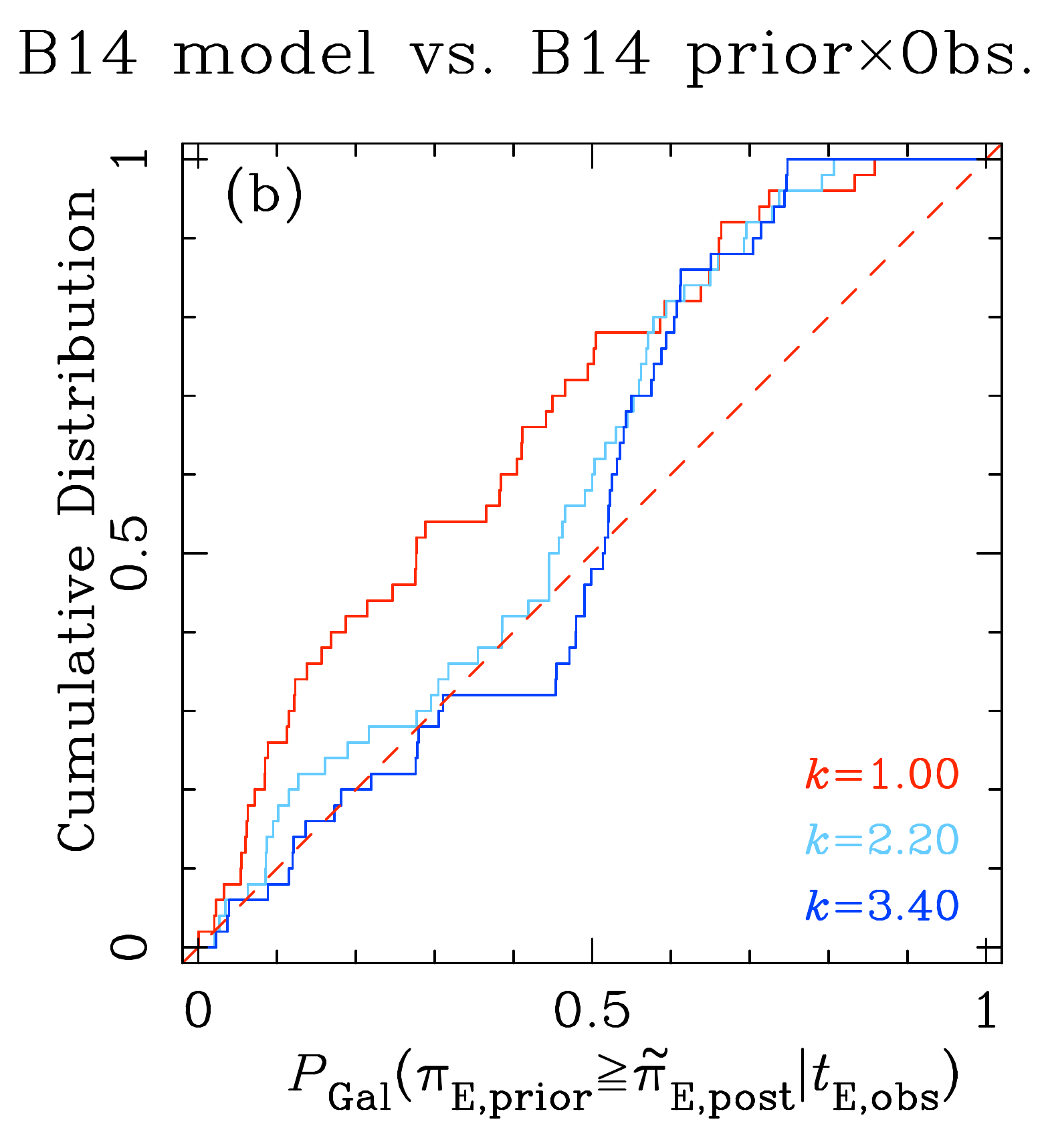}
\end{minipage}
\caption{ Panel (a) compares of the {\it Spitzer} $\pi_{\rm E}$ measurements to the B14 Galactic model, 
similar to Figure \ref{fig-tEvtil}(c). The red circles and blue squares indicate the median values, 
$\widetilde{ \pi}_{\rm E, post}$, of the posterior distributions, $f_{\rm post} (\pi_{\rm E})$,  
with the $\piEbold$ error bar inflation factors $k = 1.0$ and $k = 3.4$, respectively.
The red open circles correspond to the black open circles in Figure \ref{fig-tEvtil}(c).
Panel (b) shows the cumulative distribution of the inverse percentiles corresponding 
to the median $\widetilde{ \pi}_{\rm E, post}$, $P_{\rm Gal}  (\pi_{\rm E, prior} \geq \widetilde{ \pi}_{\rm E, post} \, | \,  t_{\rm E, obs})$, with error bar inflation
factors of $k = 1.0$ (red), $k = 2.2$ (cyan), and $k = 3.4$ (blue). }
\label{fig-B14_NF}
\end{figure}

One issue that we face is the choice of the measured value to be used for $\pi_{\rm E}$ in equation (\ref{eq-Pgal}). 
Our aim is to compare the posterior distributions, $f_{\rm post} (\pi_{\rm E})$, for the measured values with the 
Galactic model prior distributions $\Gamma_{\rm Gal} (\pi_{\rm E} \, | \,  t_{\rm E, obs})$, and our first choice is to
use the median, $\widetilde{ \pi}_{\rm E, post}$, to represent the posterior distribution ($f_{\rm post} (\pi_{\rm E})$).
This is the same choice that we made for Figure~\ref{fig-tEvtil}, but now we are preparing to invoke this choice in
a more elaborate statistical test. This more elaborate test will also involve the error bar inflation factors $k$ that were
first discussed in the contest of Figure~\ref{fig-piE2d}.
We 
also consider $k > 1$ because the $f_{\rm post} (\pi_{\rm E})$ distributions with $k = 1.0$ appear to be 
inconsistent with the Galactic model according to our visual inspection on Figure~\ref{fig-tEvtil} and the binomial tests discussed above.
Figure~\ref{fig-B14_NF}(a) shows the effect on $\widetilde{ \pi}_{\rm E, post}$ values by
applying an error bar inflation factor of 3.4 to the posterior, $f_{\rm post} (\pi_{\rm E})$, distribution
for model B14.

Figure~\ref{fig-B14_NF}(b) shows the cumulative distribution of the inverse percentiles calculated for the median, $\widetilde{ \pi}_{\rm E, post}$, i.e., 
the cumulative $P_{\rm Gal}  (\pi_{\rm E, prior} \geq \widetilde{ \pi}_{\rm E, post} \, | \,  t_{\rm E, obs})$ distribution
for three different error bar inflation factors, $k = 1.0, 2.2$, and 3.4. 
For $k = 1.0$, in red, the cumulative distribution is very far from the expected uniform distribution 
(red dashed line), as expected from the binomial tests discussed above.
But for $k = 2.2$, the cumulative inverse percentile distribution (cyan) for the 
$\widetilde{ \pi}_{\rm E, post}$ values comes close to the uniform distribution for 
$P_{\rm Gal}  (\pi_{\rm E, prior} \geq \widetilde{ \pi}_{\rm E, post} \, | \,  t_{\rm E, obs}) < 0.5$,
but it rises up to meet the $k=1.0$ solution for 
$P_{\rm Gal}  (\pi_{\rm E, prior} \geq \widetilde{ \pi}_{\rm E, post} \, | \,  t_{\rm E, obs}) > 0.6$. 
For an error bar inflation value of
$k = 3.4$, the effect is even more extreme, and we begin to see a collection of events at
$P_{\rm Gal}  (\pi_{\rm E, prior} \geq \widetilde{ \pi}_{\rm E, post} \, | \,  t_{\rm E, obs}) \approx 0.5$. 
This is also seen in Figure~\ref{fig-B14_NF}(a) where there is
a collection of events near the median (black solid line).  This is not surprising. 
In the limit where $k\rightarrow \infty$, the $f_{\rm post} (\pi_{\rm E})$ values will approach
the prior values, since a measurement with infinite error bars is equivalent to no measurement at all. Thus, the
$\widetilde{ \pi}_{\rm E, post}$ values will approach the medians of the prior
distribution and the cumulative distribution will approach a step function at 
$P_{\rm Gal}  (\pi_{\rm E, prior} \geq \widetilde{ \pi}_{\rm E, post} \, | \,  t_{\rm E, obs}) = 0.5$.

Statistically, the application of $k > 1.0$ should generate posterior distributions $f_{\rm post} (\pi_{\rm E})$
that are a better match to the models. 
With no error bar inflation,
the Bayesian posterior median values are above the median
B14 model ($\pi_{\rm E} = \widetilde{ \pi}_{\rm E, B14}$) values for 37 of 50 events, and the probability that 
$\widetilde{ \pi}_{\rm E, post} > \widetilde{ \pi}_{\rm E, B14}$ for $\geq 37$ events is $4.6\times 10^{-4}$
if we assume that the binomial distribution is appropriate.
For error bar inflation factors of $k = 2.2$ and 3.4, we have $\widetilde{ \pi}_{\rm E, post} > \widetilde{ \pi}_{\rm E, B14}$ for
30 and 24 events, respectively. The binomial distribution would imply probabilities of 0.10 and 0.66 for 
these two cases of $k = 2.2$ and 3.4, respectively. We can also apply the Anderson-Darling test formula
to get ``probabilities" of $1.7 \times 10^{-5}$, $8.4 \times 10^{-5}$, and $9.1 \times 10^{-3}$ for $k = 1.0$,
$k = 2.2$ and 3.4, respectively. 

However, when the error bars are large, the conditions for the AD test do not really apply. The test was
designed to compare a theoretical distribution with measurements without regard to error bars. In a situation
with large error bars and a steeply varying prior, the posterior distributions of the measurements can be
significantly different from the measured values. As a result, we can see a significant shift of the 
$\widetilde{ \pi}_{\rm E, post}$ values toward the median of the prior distribution, as we have see from 
Figure~\ref{fig-B14_NF} for $k=3.4$.

\begin{figure}
\begin{minipage}{0.45\hsize}
\centering
\includegraphics[width=82mm]{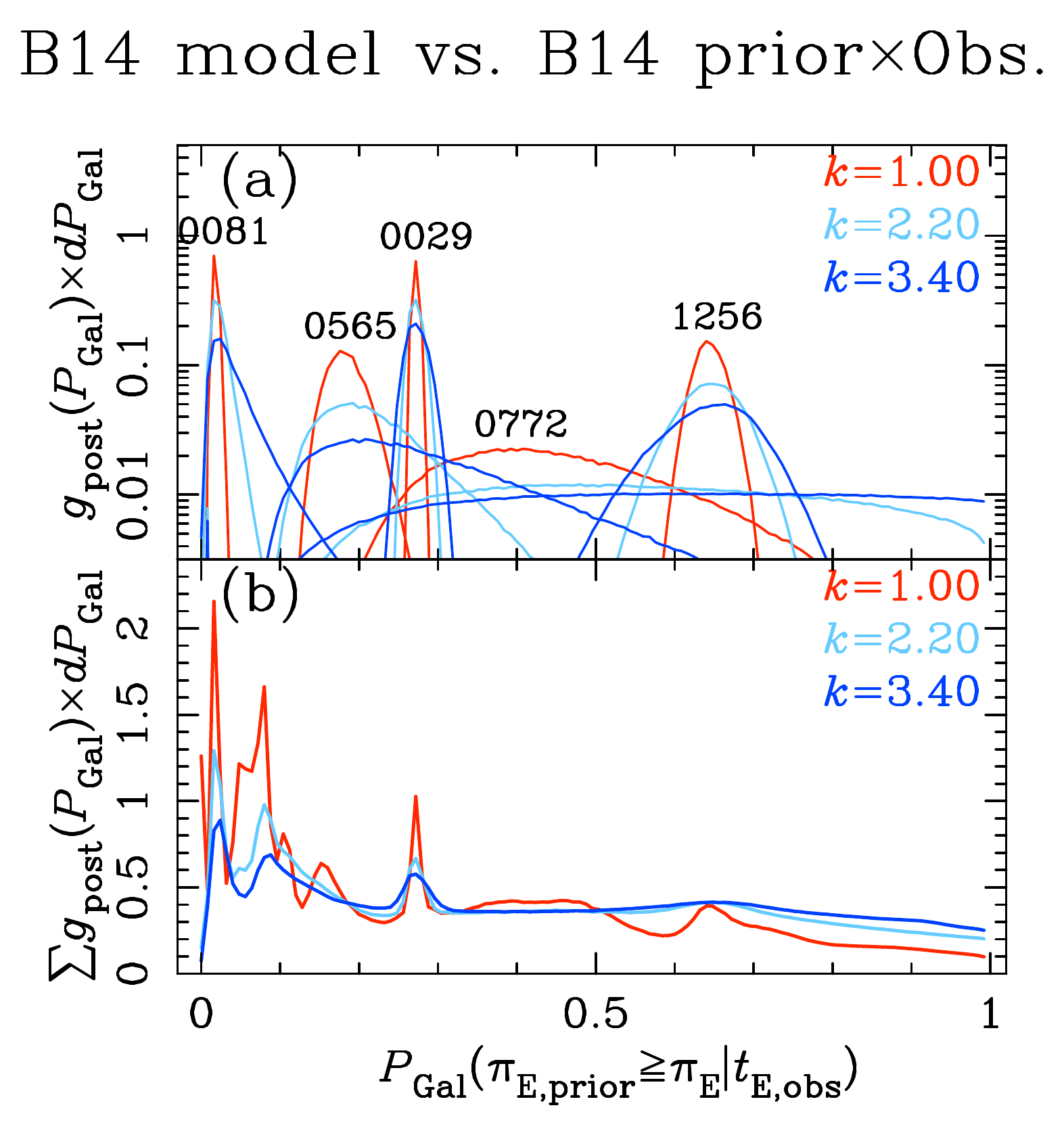}
\end{minipage}
\begin{minipage}{0.55\hsize}
\centering
\includegraphics[width=95mm]{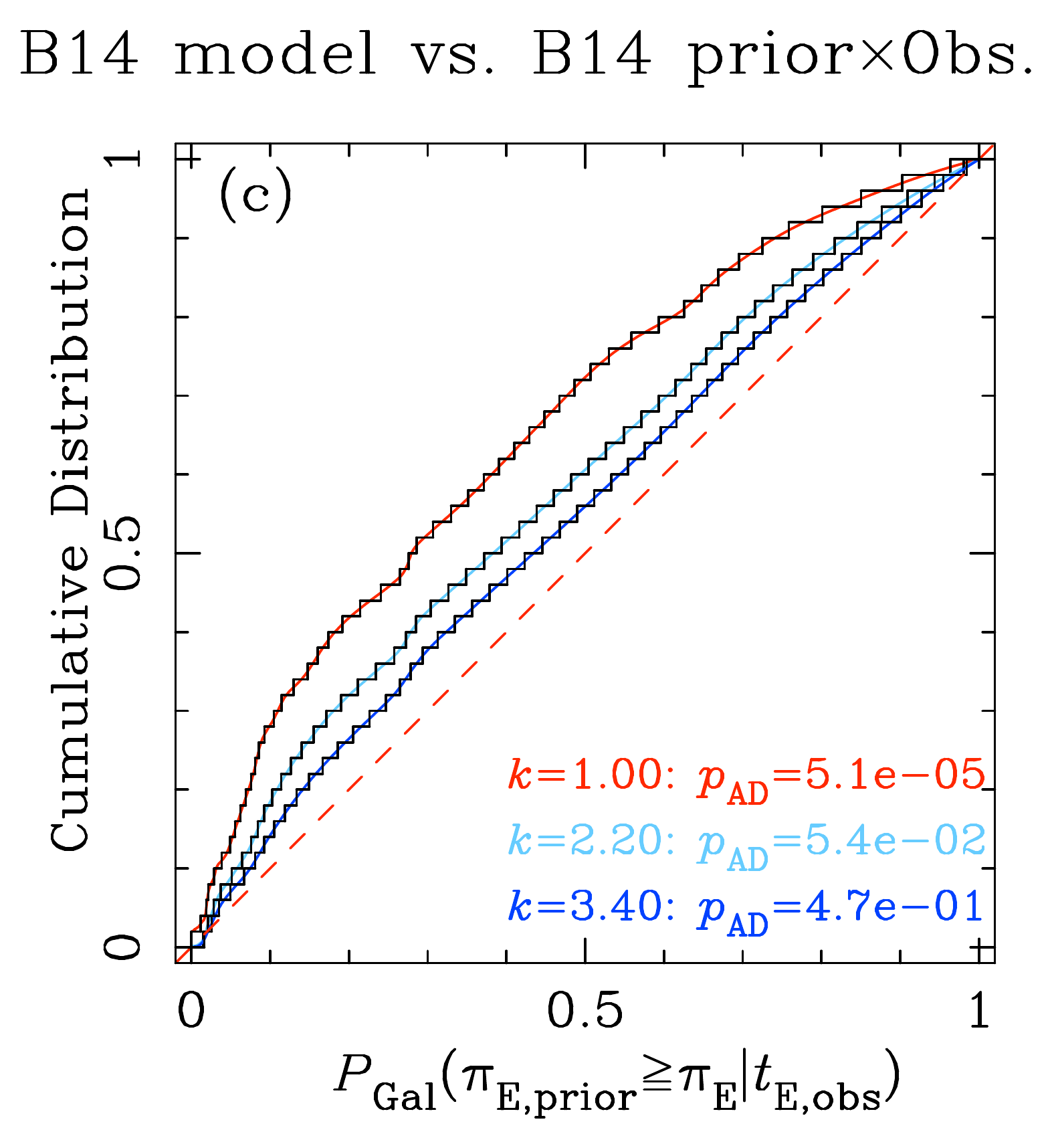}
\end{minipage}
\caption{
Panel (a) shows the posterior probability distributions of the inverse percentile
$P_{\rm Gal} (\pi_{\rm E, prior} \geq \pi_{\rm E} \, | \,  t_{\rm E, obs})$,
$g_{\rm post} (P_{\rm Gal})$, for 5 events from the
\citet{zhu17} sample with the B14 Galactic prior.
The y-axis values are probabilities integrated over bins of width $dP_{\rm Gal} =  0.008$. 
The red, cyan, and blue curves show the probability distributions as the error bar inflation factors change 
from $k = 1.0$ (red), to $k = 2.2$ (cyan), and finally to $k = 3.4$ (blue). Panel (b) shows the sum of 
these distributions over all 50 events of the \citet{zhu17} sample,
$\sum g_{\rm post} (P_{\rm Gal})$, with the same 
color scheme for the error bar inflation factors.
The cumulative density functions (CDFs) of the inverse percentile, $G_{\rm post} (P_{\rm Gal})$,
are shown in the panel (c).
The over-plotted black discrete cumulative distributions are created by dividing the vertical axis 
into 50 (i.e., number of events) bins, and
these are used for our statistical tests. The results of AD tests are shown in the bottom right of panel (c).
}
\label{fig-B14_NF2}
\end{figure}

The simplest way to modify the cumulative distributions in Figure~\ref{fig-B14_NF}(b)
to allow proper AD tests would be to randomly sample $\pi_{\rm E}$ 
from the posterior distribution $f_{\rm post} (\pi_{\rm E})$ for each event to obtain 
a sample of randomly selected ``measurements" that can be used to create inverse percentile distributions that 
can be used for the AD tests. We have tried this, but the results are somewhat noisy due to the limited
number of events. An alternative, but less noisy approach is to convert 
$f_{\rm post} (\pi_{\rm E})$ into the posterior distribution of the inverse percentile
$P_{\rm Gal} (\pi_{\rm E, prior} \geq \pi_{\rm E} \, | \,  t_{\rm E, obs})$, which we refer to as
$g_{\rm post} (P_{\rm Gal} (\pi_{\rm E, prior} \geq \pi_{\rm E} \, | \,  t_{\rm E, obs}))$. 
The relation between the two posterior distributions is
\begin{equation}
g_{\rm post} (P_{\rm Gal} (\pi_{\rm E, prior} \geq \pi_{\rm E} \, | \,  t_{\rm E, obs})) \, dP_{\rm Gal} =  f_{\rm post} (\pi_{\rm E}) \, d\pi_{\rm E} \, ,
\end{equation}
where $dP_{\rm Gal}$ is the differential of the variable 
$P_{\rm Gal} (\pi_{\rm E, prior} \geq \pi_{\rm E} \, | \,  t_{\rm E, obs})$.
The posterior inverse percentile distribution 
$g_{\rm post} (P_{\rm Gal} (\pi_{\rm E, prior} \geq \pi_{\rm E} \, | \,  t_{\rm E, obs}))$
is easily calculated by converting each $\pi_{\rm E}$ which follows the posterior distribution
$f_{\rm post} (\pi_{\rm E})$ into the inverse percentile, 
$P_{\rm Gal} (\pi_{\rm E, prior} \geq \pi_{\rm E} \, | \,  t_{\rm E, obs})$, using Equation (\ref{eq-Pgal}).
Hereafter, we use the shorter notation $g_{\rm post} (P_{\rm Gal})$ in place of
$g_{\rm post} (P_{\rm Gal} (\pi_{\rm E, prior} \geq \pi_{\rm E} \, | \,  t_{\rm E, obs}))$.
After $g_{\rm post} (P_{\rm Gal})$ is calculated for every event, we sum these distributions for all 
50 events and we refer to this combined 
distribution as $\sum g_{{\rm post}} (P_{\rm Gal})$, or $\sum_{i = 1}^{N_{\rm eve}} g_{{\rm post}, i} (P_{\rm Gal})$ 
when we want to explicitly indicate the region for the summation. $(N_{\rm eve} = 50$ in this case.)
Then, we calculate the cumulative density function (CDF) of the inverse percentile
\begin{equation}
G_{\rm post} (P_{\rm Gal}) \equiv \int_{0}^{P_{\rm Gal}} \sum_{i = 1}^{N_{\rm eve}} g_{{\rm post}, i} (P_{\rm Gal}')/N_{\rm eve} \, dP_{\rm Gal}',
\end{equation}
and divide the vertical axis into $N_{\rm eve}$ (=50) bins to make a discrete cumulative distribution which is used in the AD tests.
This is demonstrated in Figure~\ref{fig-B14_NF2}(a), which shows individual 
$g_{\rm post} (P_{\rm Gal})$ for five events with the error bar inflation factors of 
$k = 1.0, 2.2,$ and 3.4 as red, cyan, and 
blue curves respectively. 
 Figure~\ref{fig-B14_NF2}(b) shows the 
sum of $g_{\rm post} (P_{\rm Gal})$ for all 50 events, $\sum g_{\rm post} (P_{\rm Gal})$, with the same 
color scheme as Figure~\ref{fig-B14_NF2}(a). Peaks due to events 0029, 0081, and 1256
are visible in this $\sum g_{\rm post} (P_{\rm Gal})$ distribution without error bar inflation, but the event 1256 peak
is washed out with an inflation factor of $k = 3.4$, while the peaks corresponding to the other events
are diminished. Figure~\ref{fig-B14_NF2}(c) shows the CDF of the inverse percentile, $G_{\rm post} (P_{\rm Gal})$, using 
the three error bar inflation factors $k = 1.0, 2.2,$ and 3.4 in red, cyan and blue. The AD tests give
probabilities of $p_{\rm AD} = 5.1\times 10^{-5}, 0.054$, and 0.47. This confirms that the posterior 
distribution with $k = 3.4$ is consistent with the Galactic model, while the posterior 
distribution with $k = 2.2$ is only marginally consistent with the Galactic model, while the
results with the reported error bars are inconsistent with the Galactic model.

\begin{figure}
\centering
\includegraphics[width=130mm]{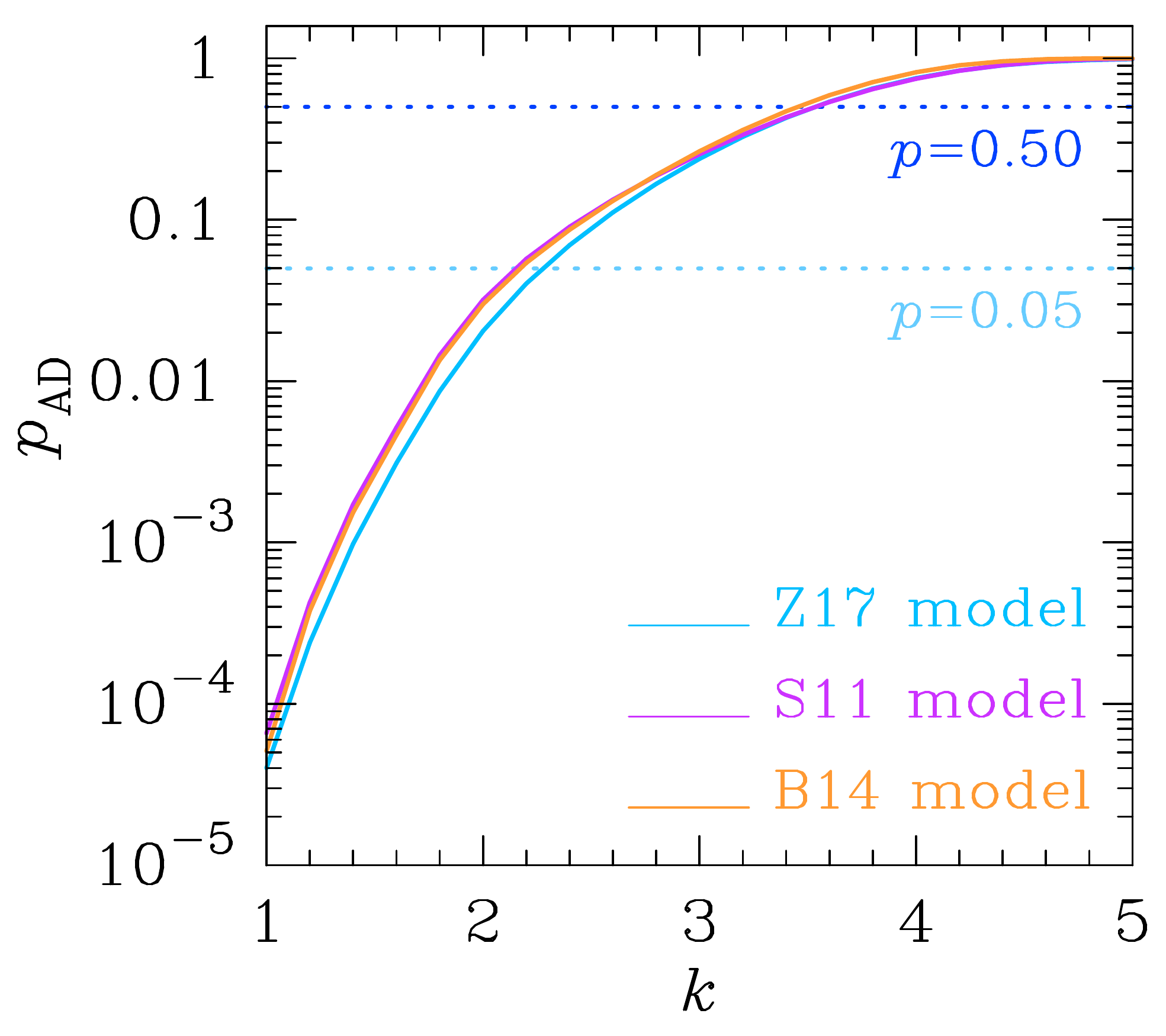}
\caption{The $p$-values of AD tests for the CDF $G_{\rm post} (P_{\rm Gal})$ 
 as a function of the error bar inflation factor, $k$ for each of our three models, Z17, S11, and B14.}
\label{fig-k-p}
\end{figure}

We repeat these calculations with all three Galactic models to determine the probabilities with $k = 1$ in addition to the minimum value 
of $k$ needed to be marginally consistent ($p_{\rm AD} = 0.05$) and fully consistent ($p_{\rm AD} = 0.50$)
with each of the Z17, S11, and B14 Galactic models, as indicated in Table~\ref{tab-results}
and Figure~\ref{fig-k-p}. 
Table~\ref{tab-results} shows that the posterior distribution with any Galactic prior fails to match the model predictions with
AD probabilities ranging from $4.0\times 10^{-5}$ to $6.6\times 10^{-5}$ when $k = 1$.
Thus, there is a strong and obvious contradiction
between the {\it Spitzer} microlensing parallax results and our Galactic models.
It is notable that one of our Galactic models is the Z17 model presented by the {\it Spitzer} microlensing
team \citep{zhu17}, but the Z17 model does not fit the distribution of {\it Spitzer} microlensing parallax results
significantly better than the other two models. 
We find that $k > 2.2$ is minimum error bar correction needed for each of the north and 
east components of the reported $\piEboldobs$ measurements to make the {\it Spitzer} $\pi_{\rm E}$ 
measurements consistent with the Galactic models, but an error bar inflation factor of $k = 3.4$ or 3.5
is needed to match the median prediction of the Galactic models.

\begin{deluxetable}{cccccc}
\tablecaption{Results of AD tests to compare the posterior distributions for the 50 events in the Z17 raw sample and the Galactic model. \label{tab-results}}
\tablehead{
\colhead{Model} & \colhead{$A^2$ ($p_{\rm AD}$) w/ $k=1$} & \colhead{$k\,(p_{\rm AD} = 0.05)$\tablenotemark{a}} & \colhead{$k\,(p_{\rm AD} = 0.50)$\tablenotemark{b}} & \colhead{$n_{\rm D/B, th}$\tablenotemark{c}}
}
\startdata
Z17 &  9.01 ($4.0 \times 10^{-5}$)  & 2.27 & 3.53 & 3.1 \\
S11 &  8.54 ($6.6 \times 10^{-5}$)  & 2.14 & 3.53 & 3.9 \\
B14 &  8.78 ($5.1 \times 10^{-5}$)  & 2.17 & 3.45 & 4.0 \\
\enddata
\tablenotetext{a}{Minimum value of error inflation factor $k$ to be $p_{\rm AD} > 0.05$.}
\tablenotetext{b}{Minimum value of error inflation factor $k$ to be $p_{\rm AD} > 0.50$.}
\tablenotetext{c}{Minimum $n_{\rm D/B}$ value to be $p_{\rm AD} > 0.05$ within $0.2 < \alpha_{\rm bd} < 1.3$ when $k = 1$.}
\tablecomments{Indicated model is used both for the Bayesian prior convolved with the Z17 measurements and for the compared model.}
\end{deluxetable}

We have confirmed our method of using $G_{\rm post} (P_{\rm Gal})$ for AD tests 
by conducting simple AD tests using $\pi_{\rm E}$ values randomly sampled from 
$f_{\rm post} (\pi_{\rm E})$ for each event.
From 600 trials using the B14 model, we find median values of 
$p_{\rm AD} = 5.2\times 10^{-5}$, $0.042$ and 0.30, for $k = 1.0, 2.2$ and 3.4, respectively.
These are very close to the $p_{\rm AD}$ values of $5.1\times 10^{-5}, 0.054$, and 0.47 that
we found using $G_{\rm post} (P_{\rm Gal})$.
The $1~\sigma$ ranges of these  $p_{\rm AD}$ values are noisy, as expected. They are
$5.8\times 10^{-6}$ - $3.0\times 10^{-4}$, $6.9\times 10^{-3}$ - $0.17$ and $0.069$ - $0.70$, respectively.

One might think that the AD test is not designed to be part of a Bayesian analysis. 
We use the posterior distributions in this paper, though, because we want to be conservative by adding the Galactic prior.
We find $p_{\rm AD}$ values of $9.1 \times 10^{-9}$, $3.0 \times 10^{-8}$ and $6.1 \times 10^{-9}$ with the Z17, S11 and B14 model, respectively, when 
we conduct the AD tests using the original reported $\pi_{\rm E}$ values by \citet{zhu17}.
In the tests, we use a solution that has minimum $\chi^2 + 4 \ln \pi_{\rm E}$ value for each event. These values are smaller than the probabilities with
 the posterior distribution in Table \ref{tab-results} by 3-4 orders of magnitude and this confirms that we are conservative.

\section{What Is the Cause of the Discrepancy?}\label{sec-discrep}

The most obvious potential cause of this discrepancy between the three Galactic models we 
consider and the {\it Spitzer} microlensing parallax values is systematic errors in the {\it Spitzer} photometry.
This is why we considered the error bar inflation factor $k > 1$ in Section \ref{sec-test}.
Section 5.1 of \citet{zhu17} is devoted to a discussion of systematic errors in the {\it Spitzer} photometry, and they
mention five events with ``prominent" deviations of the {\it Spitzer} photometry from the best fit light curve.
Our visual inspection of the 50 light curves presented in \citet{zhu17} indicates that 18 (or 36\%) of these have obvious
systematic differences between the {\it Spitzer} photometry and the best fit microlensing models. Since the 
microlensing parallax parameters are often determined almost solely from the {\it Spitzer} data, it seems
quite plausible that some of the events without obvious systematic photometry errors may, nevertheless, have large
errors that can be accounted for with incorrect $\piEboldobs$ model parameters. \citet{zhu17} suggest that these 
systematic errors might not cause problems with the {\it Spitzer}  $\piEboldobs$ measurements with a reference to
three events \citep{uda15,pol16,han17} for which there is some evidence suggest that the systematic errors
may not have much influence on the  $\piEboldobs$. But, two of these events have much stronger microlensing
signals in the  {\it Spitzer} data than is typical, so this argument may not apply to the bulk of the \citet{zhu17}
sample.

Although these obvious systematic photometry problems are an important issue, we also need to 
consider several other issues that could contribute to this discrepancy. We start by considering the
possibility that a plausible Galactic model could be consistent with the data.
Then we consider possibility of bias in the {\it Spitzer} sample could influence the large $\pi_{\rm E}$ values.
We set the error bar inflation factor $k = 1$ in all of our subsequent analysis.

\subsection{Is There a Galactic Model That Can Match the Spitzer Data?}\label{sec-artest}

In this section, we consider modifications to our Galactic models to match the distribution of
$\piEboldobs$ values from the \citet{zhu17} {\it Spitzer} sample. For fixed $t_{\rm E}$ values, microlensing parallax 
values can be increased by decreasing the lens distance, $D_L$, the lens mass, $M_L$, and/or the lens-source
relative transverse velocities. However, the requirement of fixed $t_{\rm E}$ values, means that the distributions
of $D_L$, $M_L$, and transverse velocity are correlated, and in this section, we consider modifications of  
the $M_L$ and $D_L$ distributions of the Galactic models. One parameter that is not very well known is  
the slope of the initial mass function (IMF) in the brown dwarf mass region, $\alpha_{\rm bd}$.
This has been measured in previous microlensing studies \citep{sum11,mro17}, but these measurements 
depend on Galactic models similar to (or identical to) the models that we have considered. So, it is sensible to
consider variations in the $\alpha_{\rm bd}$ values. 

The other parameter that we consider modifications to is the ratio of the disk mass to bulge mass. We define the
change relative to the fiducial model disk/bulge mass ratio as
\begin{equation}
n_{\rm D/B} \equiv \frac{(\rho_{\rm D, 0}/\rho_{\rm B, 0})_{\rm art}}{(\rho_{\rm D, 0}/\rho_{\rm B, 0})_{\rm org}}, \label{eq-ndb}
\end{equation}
where $(\rho_{\rm D, 0}/\rho_{\rm B, 0})_{\rm art}$ is the value in an ``artificial" model with an increased
disk mass and/or a decreased bulge mass, and $(\rho_{\rm D, 0}/\rho_{\rm B, 0})_{\rm org}$ is the value in the unmodified models presented in 
Table \ref{tab-models}.

\begin{figure}
\begin{minipage}{0.32\hsize}
\centering
\includegraphics[width=60mm]{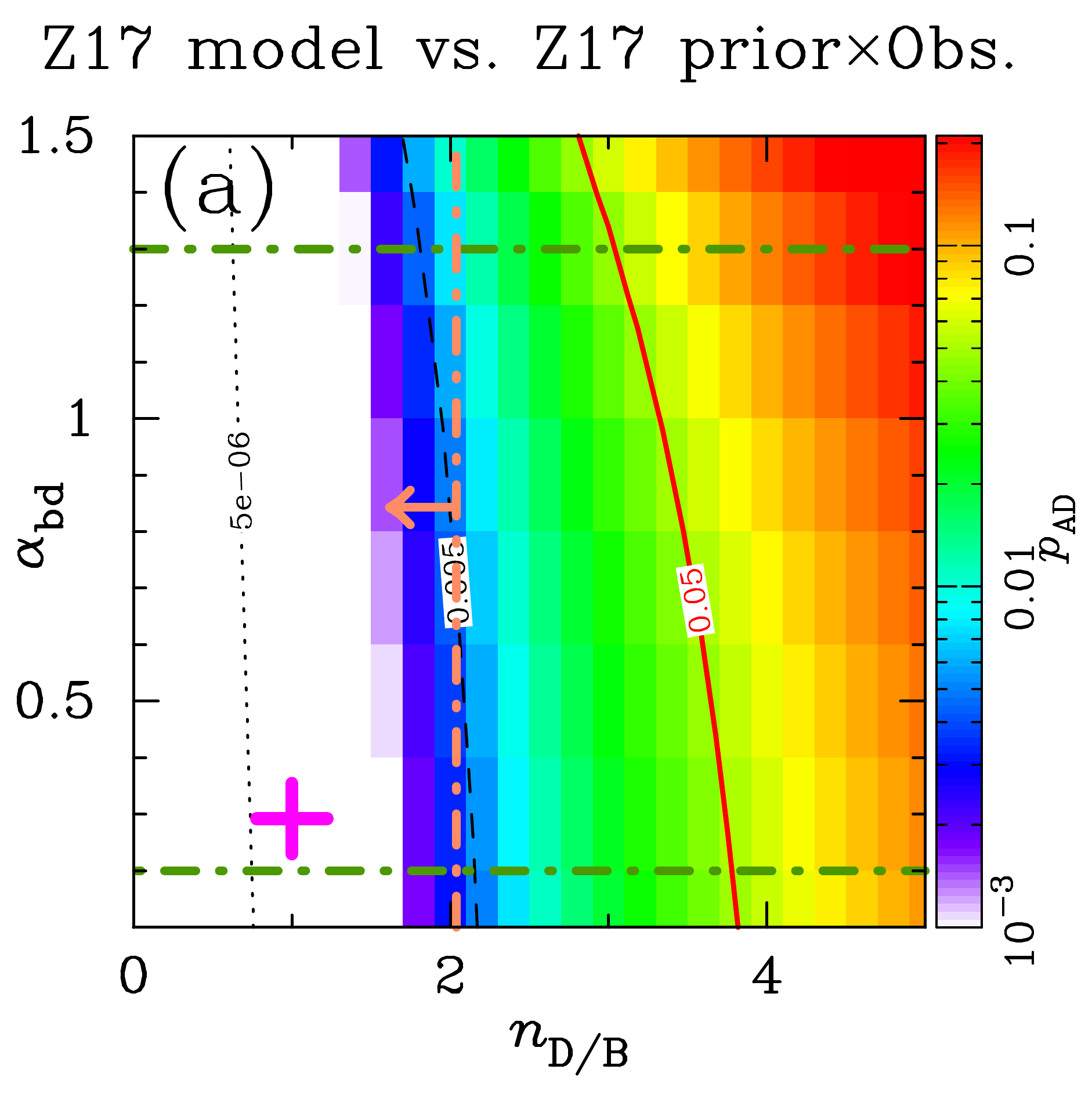}
\end{minipage}
\begin{minipage}{0.32\hsize}
\centering
\includegraphics[width=60mm]{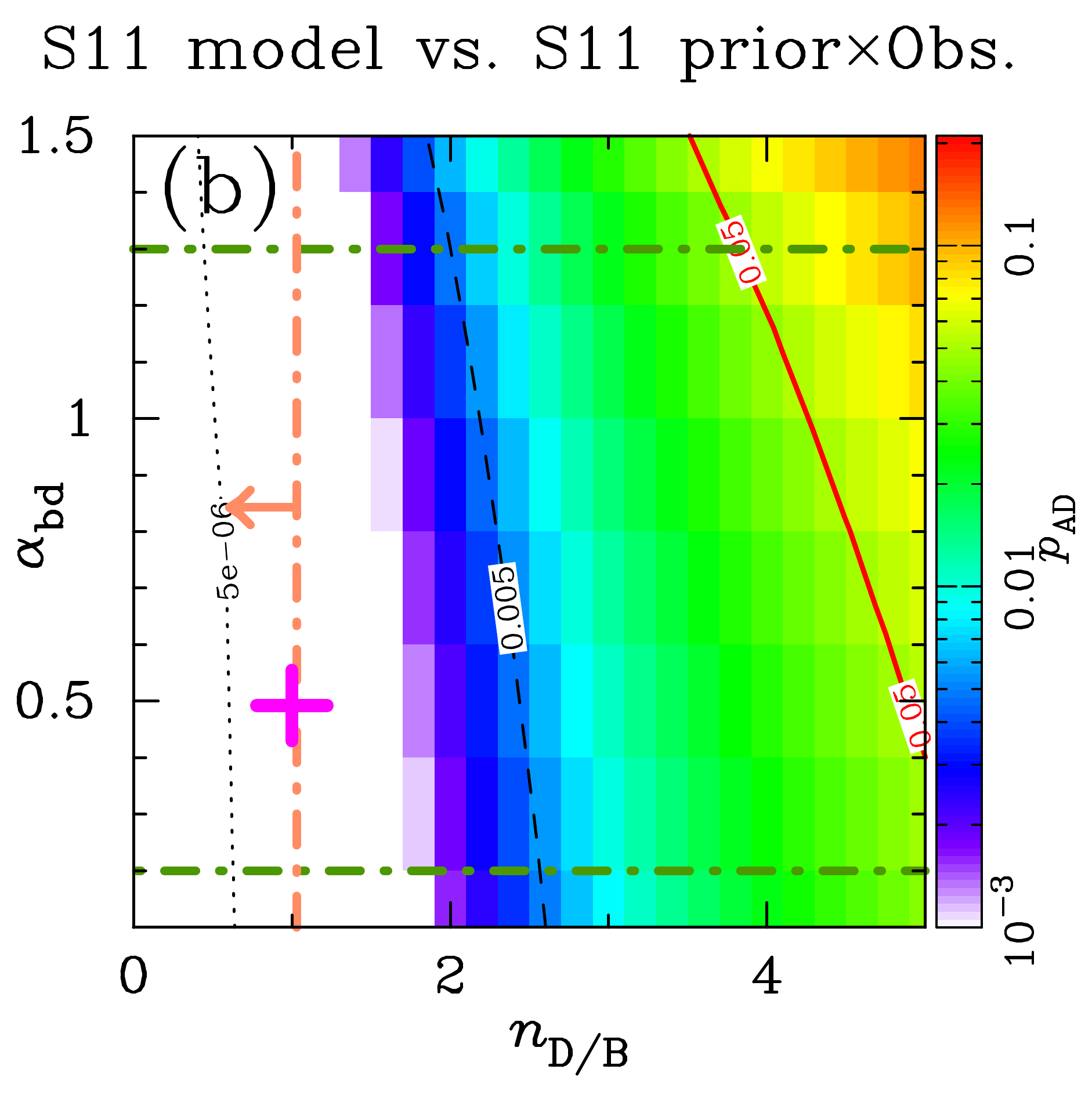}
\end{minipage}
\begin{minipage}{0.32\hsize}
\centering
\includegraphics[width=60mm]{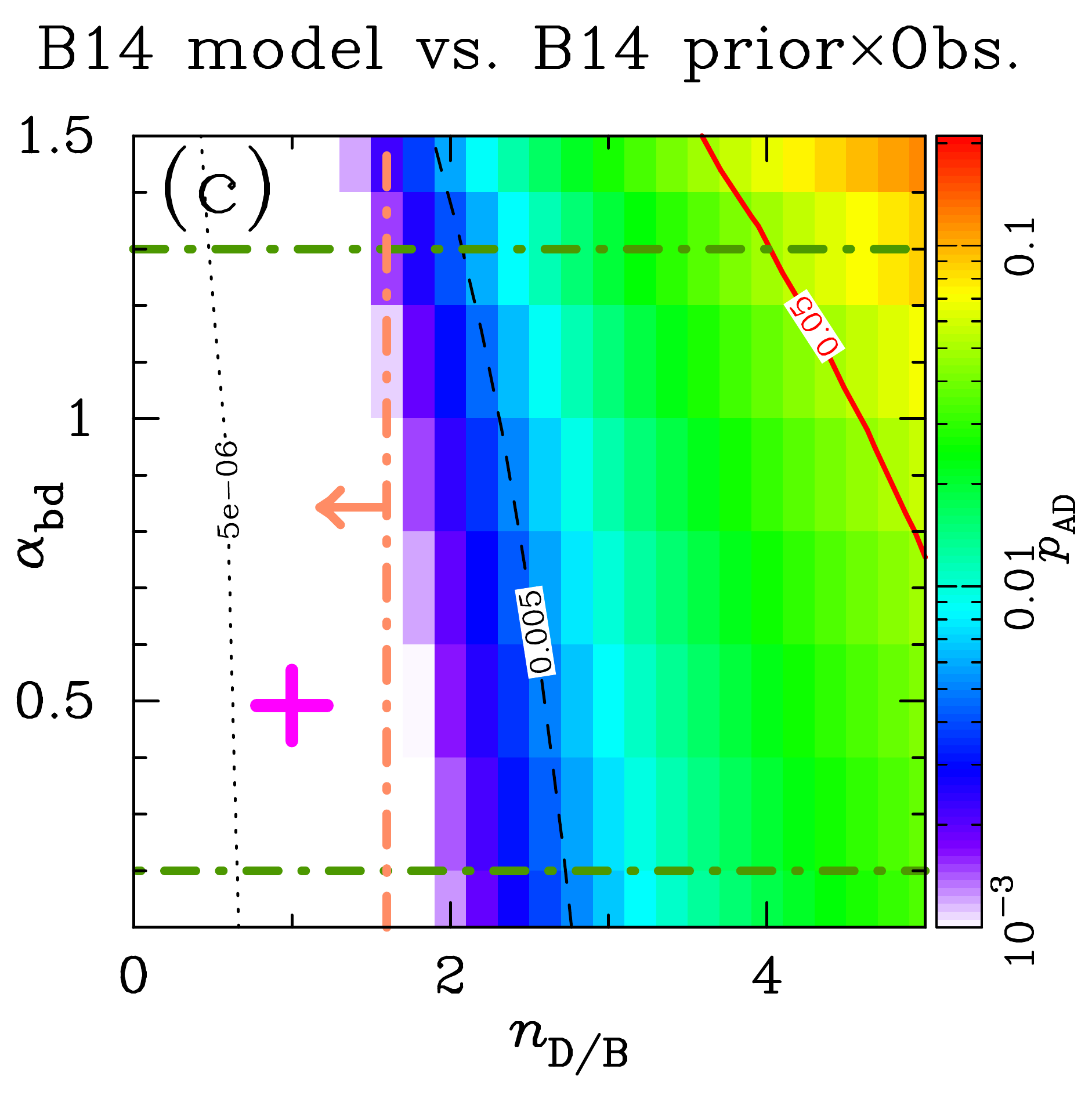}
\end{minipage}
\caption{Results of AD tests with various $(n_{\rm D/B}, \alpha_{\rm bd})$ combinations, where 
the modified Galactic model is applied both to the Bayesian prior and to the compared model. Both color 
maps and contours show $p_{\rm AD}$-values from the AD tests.
In each panel, the two horizontal dark green dashed-dotted lines indicate the $3\sigma$ range for 
$\alpha_{\rm bd}$ from \citet{mro17}, the vertical orange dashed-dotted line indicates the $3\sigma$ upper 
limit on $n_{\rm D/B}$ described in Section \ref{sec-artest}, and the magenta cross indicates the values
of  $\alpha_{\rm bd}$ and $n_{\rm D/B}$ employed by each of the three models: Z17, S11, and B14.}
\label{fig-pAD}
\end{figure}

We compare the CDFs of the inverse percentile, $G_{\rm post} (P_{\rm Gal})$,
to the expected uniform distribution, similar to the comparisons presented in Figure~\ref{fig-B14_NF2}, for modified
versions of the Z17, S11, and B14 on a grid of  $n_{\rm D/B}$ and $\alpha_{\rm bd}$
values. We perform the AD test on each such model, and present the resulting probabilities,  $p_{\rm AD}$, 
as a function of $n_{\rm D/B}$ and $\alpha_{\rm bd}$ in Figure~\ref{fig-pAD}, where the modified model
is also used for the Bayesian prior to calculate $f_{\rm post} (\pi_{\rm E})$ as well as the compared model at each grid. 
Both the color map and the contours indicate the $p_{\rm AD}$ distribution, and the red contour lines indicate
our threshold $p$-value of $p_{\rm AD} \geq 0.05$, which is marginally acceptable.

The
brown dwarf mass function slope, $\alpha_{\rm bd}$, has been measured using data from the MOA-II
\citep{sum11}, OGLE-III \citep{weg17}, and the OGLE-IV \citep{mro17} microlensing surveys, with results
that are very consistent with each other. OGLE-IV is the most sensitive survey, so we show their 3$\sigma$ 
limits, $0.2 < \alpha_{\rm bd} < 1.3$, as the green horizontal dot-dashed lines in Figure~\ref{fig-pAD}. 
If we restrict $\alpha_{\rm bd}$ to lie within this range, then we can use the distributions presented in 
Figure~\ref{fig-pAD}, to derive the minimum values of $n_{\rm D/B}$ required to pass
our acceptability threshold of $p_{\rm AD} \geq 0.05$. These results are given in Table \ref{tab-results}.
The fiducial model values of $n_{\rm D/B} = 1$ for each of the Z17, S11, and B14 models
are thought to explain the observed $t_{\rm E}$ distributions very well, so if we are required to select
$n_{\rm D/B} \gg 1$ in order to get plausible $p_{\rm AD}$ values, this could be taken
as an indication that the {\it Spitzer} results cannot be explained by any reasonable Galactic model. 
The results listed in Table \ref{tab-results} are 3.1, 3.9 and 4.0 for the Z17, S11, and B14 models, respectively.

It seems implausible that the disk-to-bulge mass ratio for the Z17, S11, and B14 models could really
be increased by a factor of 3.1 or more from the model values and still be consistent with observations, but
let us consider this question in more detail.  The relative disk-to-bulge mass ratio, $n_{\rm D/B}$, as defined 
in Eq. (\ref{eq-ndb}) will increase when $\rho_{\rm D, 0}$ increases, $\rho_{\rm B, 0}$ 
decreases, or both.

From Gaia DR1 \citep{gai16}, \citet{bov17} derives a local stellar density of
$0.040 \pm 0.002 ~M_{\odot} \, {\rm pc}^{-3}$ for main sequence stars. To get the total density, we must add
the density of white dwarfs,  $0.0065 ~M_{\odot} \, {\rm pc}^{-3}$ \citep{bov17}, and brown dwarfs, which account
for 4.4\% as much mass as the main sequence stars \citep{mck15}. This gives a total density 
of $0.048 \pm 0.002 ~M_{\odot} \, {\rm pc}^{-3}$. This is $1.26 \pm 0.05$ times, $0.81 \pm 0.03$ times 
and $1.24 \pm 0.05$ times larger than the $\rho_{\rm D, 0}$ values for the Z17, S11, and B14 
models. respectively.

\citet{por17} constructed a dynamical model of the bulge, with the aid of N-body simulations, and their
model is consistent with the bulge star number counts from the VVV survey  \citep{min10}  and 
spectroscopic surveys, such as BRAVA \citep{kun12}.
Their model was also confirmed to be consistent with the OGLE-II proper motion data \citep{sum04}, 
microlensing optical depth \citep{sum16}, and the $t_{\rm E}$ distribution of OGLE-III \citep{wyr15, weg17}.
They derive stellar mass traced by red clump giants observed by the near infrared surveys in a box of 
$(\pm 2.2 \times \pm 1.4 \times \pm 1.2)$ kpc around the principal axes of the bulge to be
$(1.32 \pm 0.08) \times 10^{10}~M_{\odot}$.
We integrate the bar models of Z17, S11 and B14 within the box and obtain $1.76 \times 10^{10}~M_{\odot}$ for 
the Z17 model and $1.39 \times 10^{10}~M_{\odot}$ for both the S11 and B14 models. In order to
be consistent with the \citet{por17} bulge mass, the normalization of the model bulge masses 
must be multiplied by $0.75 \pm 0.05$ times, $0.95 \pm 0.06$ times and $0.95 \pm 0.06$ times for the Z17, S11,
and B14 models, respectively. Therefore, to be consistent with the recent studies requires $n_{\rm D/B}$ 
values for Z17, S11, and B14 models to be $1.68 \pm 0.12$, $0.85 \pm 0.06$, and  $1.30 \pm 0.10$, 
respectively. The vertical orange dot-dashed lines in  Figure~\ref{fig-pAD} represent the 3$\sigma$ upper limits
on $n_{\rm D/B}$ from this calculation, and it is clear that they do not come close to the $p_{\rm AD} \geq 0.05$
contours. In the most favorable case, the Z17 model, the $p_{\rm AD} \geq 0.05$ contour is still excluded by
$>10\sigma$ in this most favorable case. 

Although we do not consider the change of kinematic property due to the system mass change in the modified models, this effect 
is not likely to fully help the situation because a large change in $n_{\rm D/B}$ value requires both dynamically hotter disk stars and dynamically cooler bulge stars.
These two effects are roughly cancelled in the $\Gamma_{\rm Gal} (\pi_{\rm E} \, | \,  t_{\rm E, obs})$ distribution.
So we conclude that the reported {\it Spitzer} $\pi_{\rm E}$ values are
not consistent with any reasonable Galactic model although the level of discrepancy might be mitigated a little by applying a modest change to the Galactic models.

\subsection{A Bias in the Spitzer Sample Is Unlikely to Cause the Discrepancy}\label{sec-gammaDs}
Another possibility is that the discrepancy between the Galactic models and the {\it Spitzer}  $\pi_{\rm E}$
values could be due to a bias in the selection of the events in the Z17 sample.
One known bias that affects $\pi_{\rm E}$ distribution in the {\it Spitzer} sample is that the {\it Spitzer}
event selection process favors larger $t_{\rm E}$ values. This bias is discussed by \citet{zhu17},
who also pointed out that the Z17 sample lacks extremely long timescale ($t_{\rm E} \simgt 100$ days).
A bias in the $t_{\rm E}$ distribution for the Z17 sample does not affect our analysis because we consider the
event rate, $\Gamma_{\rm Gal} (\pi_{\rm E} \, | \,  t_{\rm E, obs})$, as a function of $t_{\rm E, obs}$, as
discussed in Section \ref{sec-method}. 

A modest bias that does exist in the Z17 sample concerns the source brightness. According to \citet{yee15},
events with brighter sources are favored. The selection of brighter sources provides a bias in favor of
smaller source distances, $D_{\rm S}$. The Galactic models consider a source distance dependence for
the event rate that is weighted by proportional to $D_{\rm S}^{2-\gamma}$, so we can evaluate the
effect of this bias by varying $\gamma$.

However, as described in Section \ref{sec-everate} and summarized in Table \ref{tab-models}, 
we have already used three models with very different $\gamma$ values, ranging from 1.5 for the B14 model to 2.85 
for the Z17 model, but the $p$-values for these models are very similar. This is similar to the analysis of
\citet{zhu17} who found that variation of the $\gamma$ value had little effect on their results (see Appendix A of their paper).
We have also conducted the AD-test for the B14 model with $\gamma = 2.85$, instead of the original value of 1.5, 
and we find $p_{\rm AD} = 8.7 \times 10^{-5}$.
This is a slight improvement over the original $p_{\rm AD} = 5.1 \times 10^{-5}$ value for this model, so
it does account for the discrepancy. Thus, sample bias cannot explain the failure of the  {\it Spitzer} $\pi_{\rm E,obs}$
sample to match the Galactic models.

\section{Spitzer's Systematic Photometry Errors} \label{sec-dis}
In Section \ref{sec-discrep}, we have established that the systematically large $\pi_{\rm E}$ 
values found by \citet{zhu17} are
neither likely to be an artifact of inadequate Galactic models nor caused by any selection bias in their sample.
Now we consider the idea that this discrepancy is due to systematic errors in 
the \citet{zhu17} $\pi_{\rm E}$ measurements, which seems
to be the only reasonable conclusion. The $t_{\rm E, obs}$ measurements by OGLE have been 
shown to be very accurate and unbiased \citep{mro17}, but it is possible that systematic errors in
the OGLE photometry could contribute to the systematic errors in the $\pi_{\rm E}$ measurements for
some of the longer duration events. However, the main purpose of the {\it Spitzer} survey is to measure
$\pi_{\rm E}$, and the {\it Spitzer} measurements are known to dominate the $\pi_{\rm E}$ for some
long duration events \citep{pol16}.

In this section, we first report our findings that the level of discrepancy is correlated with
three different aspects in the events in the Z17 sample. These are
the source magnitude $I_{\rm S}$, the light curve peak coverage by {\it Spitzer}, and whether or not
an obvious systematic error is seen in the {\it Spitzer} light curves shown in Figure 4 of \citet{zhu17}.
This is not only useful to understand how the systematic errors affect the parallax measurements, but also another conclusive evidence that the discrepancy we have discussed 
mainly attributes to the {\it Spitzer} light curves rather than the Galactic models or our method.
Then we discuss possible ways to account for the systematic errors including some recent attempts by other authors.
Finally we compare our study with \citet{sha19} and \citet{zang19} who conducted statistical tests on much
smaller samples of {\it Spitzer} events that might plausibly be reinterpreted as tests of consistency of 
their samples with Galactic models.

\subsection{Correlation Between Systematic Errors and Event Characteristics} \label{sec-corr}

One way to help us find nature of the systematic errors is to consider possible correlations between the
apparent systematic errors and the characteristics of the events in the sample.
To address this question, we conduct the statistical tests on a number of Z17 sub-samples based on
several different event characteristics. This investigation has revealed that the $p$-values depend 
on the source star brightness, the coverage of the light curve peak in the {\it Spitzer} data, and the
presence of obvious systematic discrepancies between the {\it Spitzer} data and the best fit light curve models.
These correlations help to elucidate the nature of the systematic errors in the {\it Spitzer} data.
For the remainder of this paper, we will use the B14 model with the error bar inflation factor $k = 1$ for our tests, unless we specifically indicate
that we are using another model. Results from other models are provided in some of the tables, and our
conclusions do not depend on which Galactic model is used.

\subsubsection{Correlation with the Source Brightness}  \label{sec-corrIs}

\begin{figure}
\begin{minipage}{0.50\hsize}
\centering
\includegraphics[width=73mm]{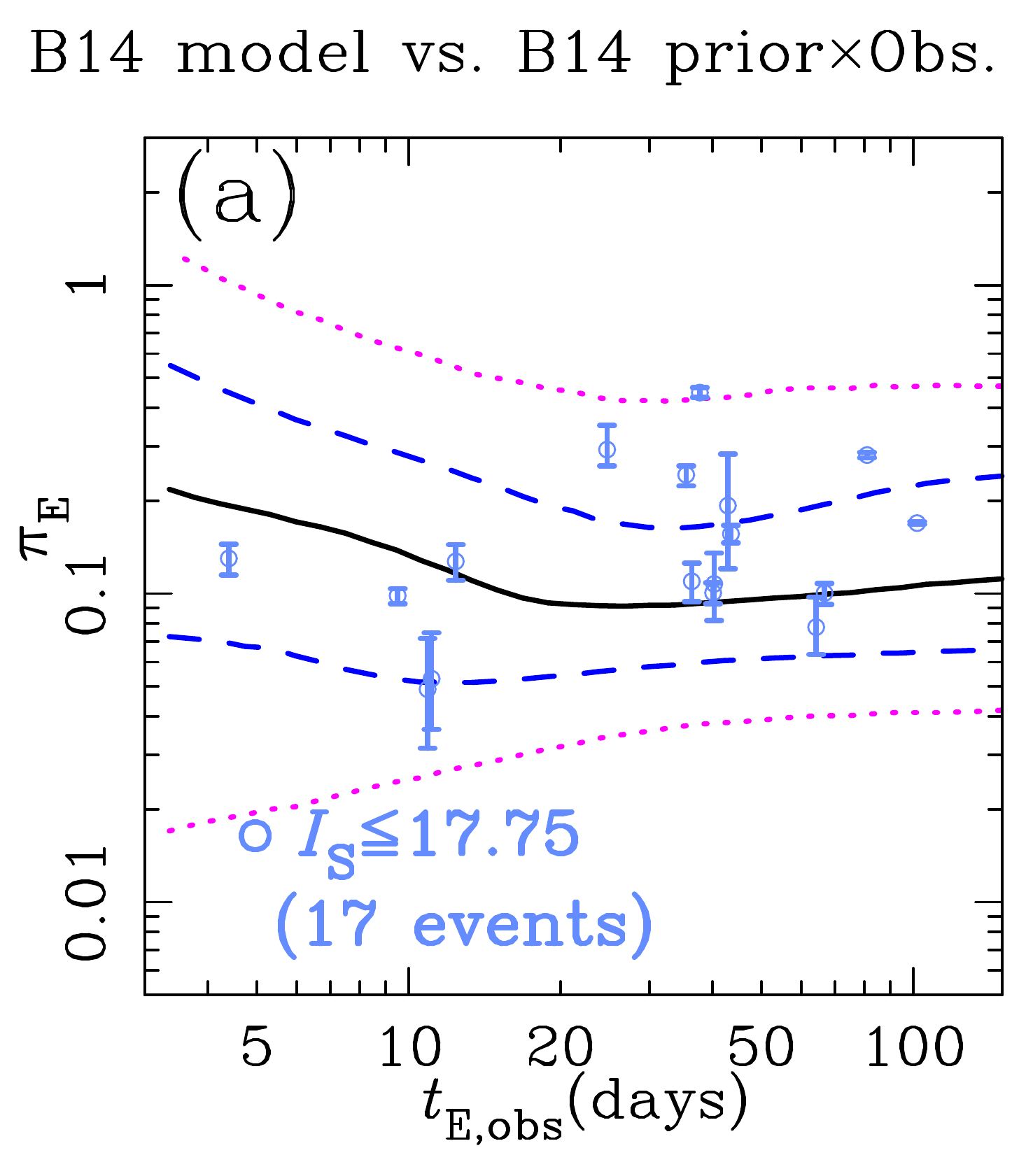}
\end{minipage}
\begin{minipage}{0.50\hsize}
\centering
\includegraphics[width=73mm]{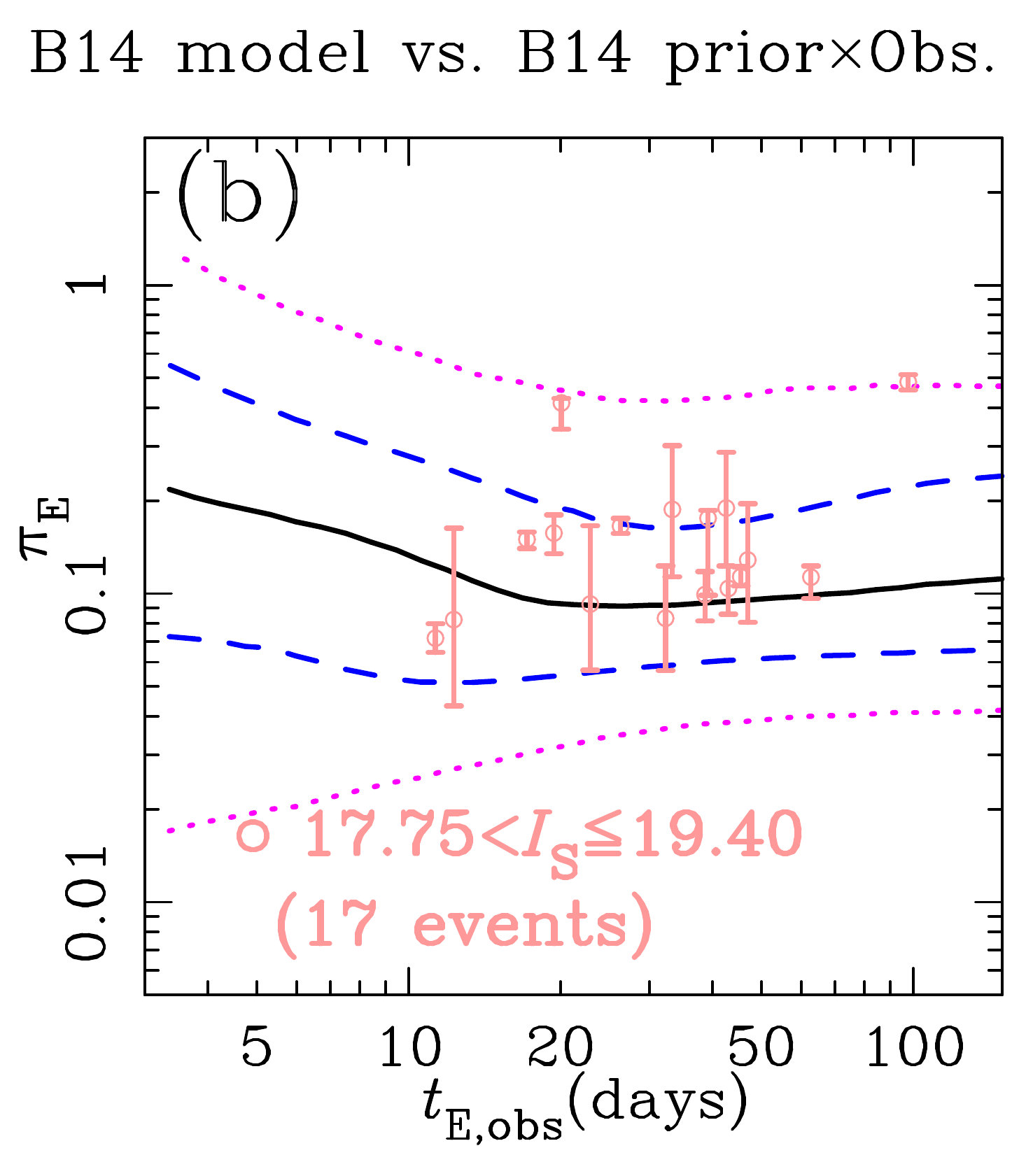}
\end{minipage}
\begin{minipage}{0.50\hsize}
\centering
\includegraphics[width=73mm]{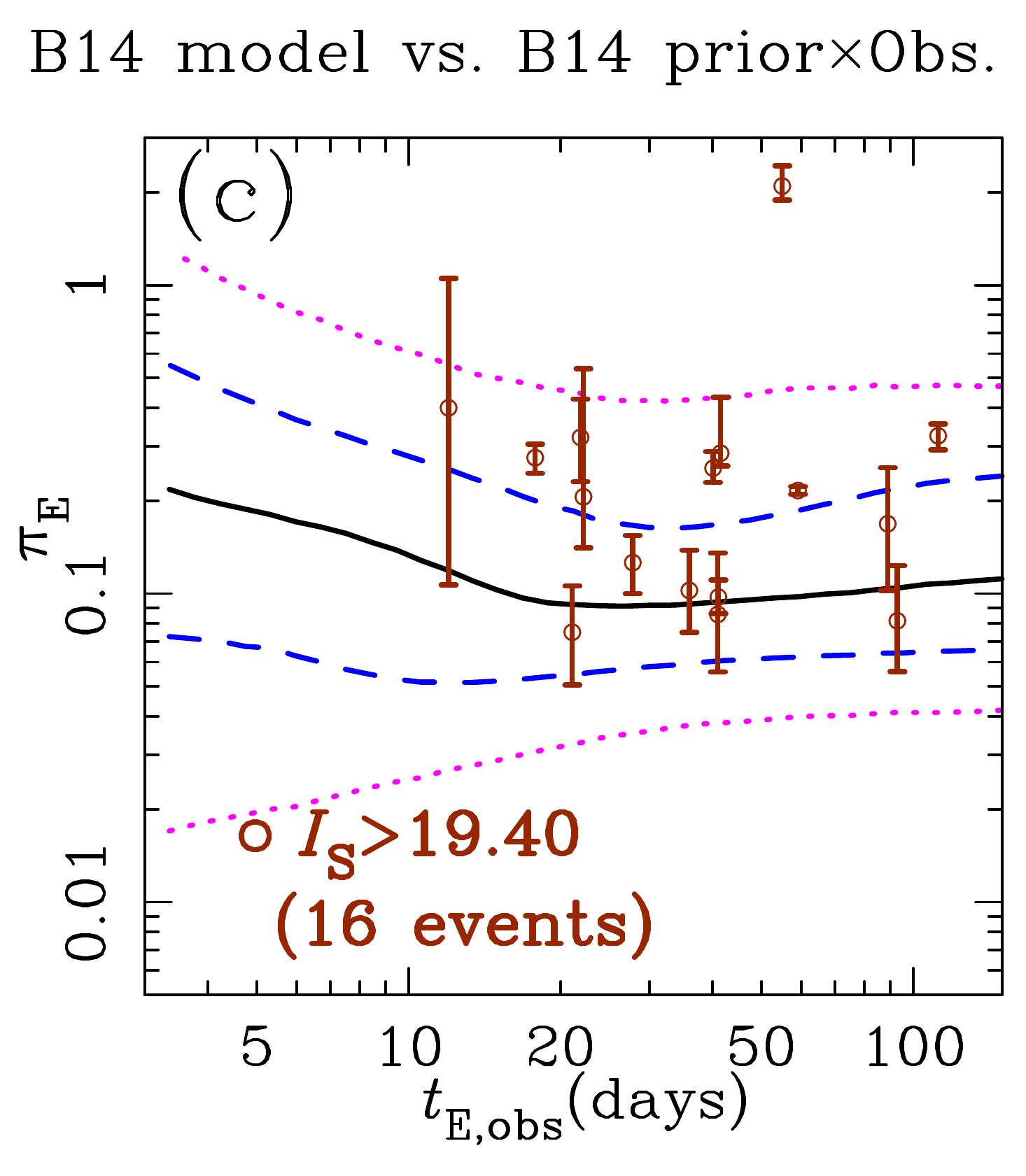}
\end{minipage}
\begin{minipage}{0.50\hsize}
\centering
\includegraphics[width=73mm]{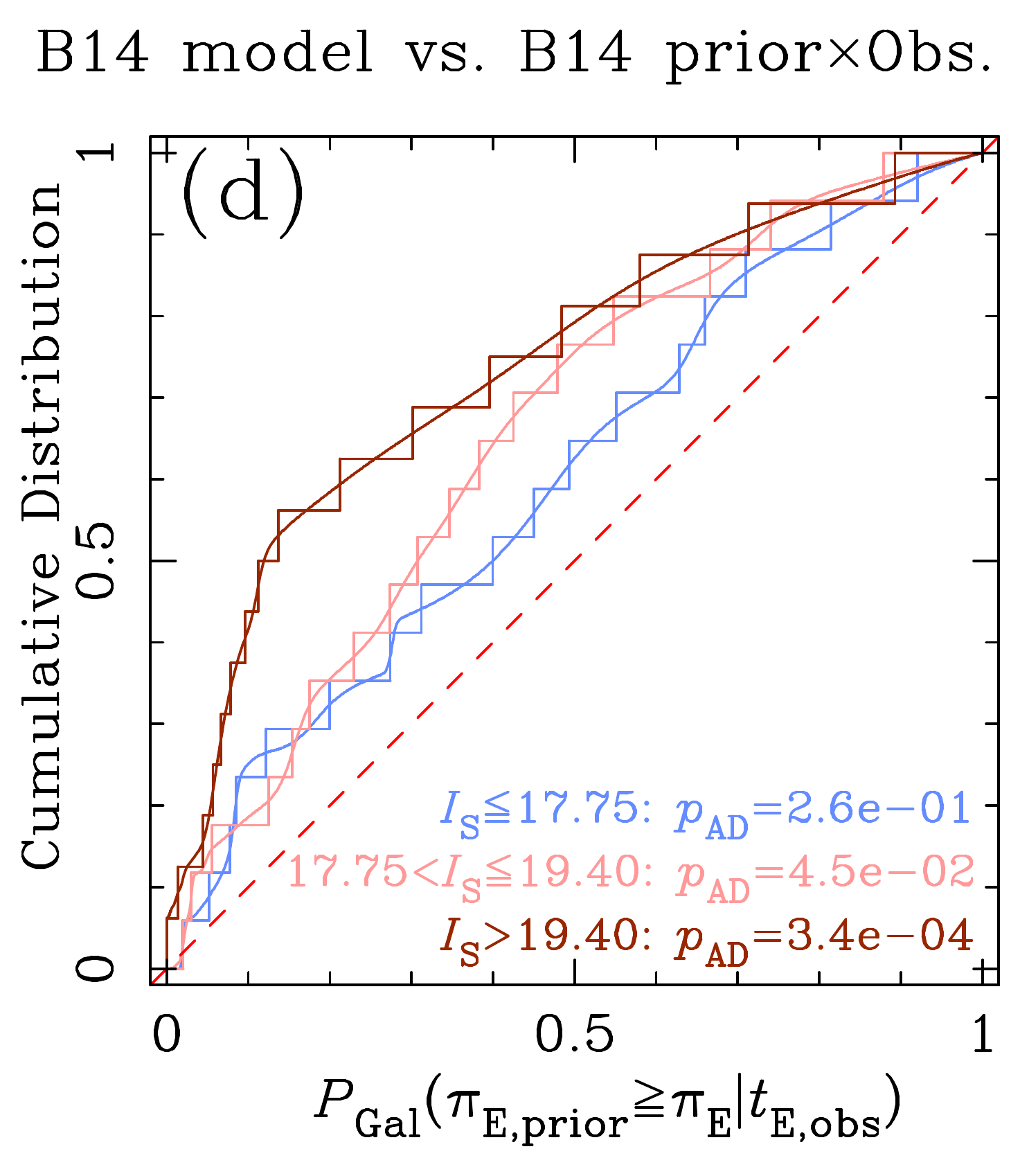}
\end{minipage}
\caption{Panels (a)-(c) are similar to Fig. \ref{fig-tEvtil} (c), but only plot events which have the source magnitude $I_{\rm S}$ in the indicated range in each panel.
Panel (d) is the CDFs of the inverse percentile, $G_{\rm post} (P_{\rm Gal})$, for each sample shown in the panels (a)-(c).}
\label{fig-3Is}
\end{figure}

In general, observations of bright stars provide a high signal-to-noise ratio and they are likely to be
less affected by systematic errors, except those, like flat-fielding errors that are proportional to the star's
brightness.
Motivated by this thought, we divide all 50 events of the \citet{zhu17} raw sample into three groups based 
on the source magnitude reported by \citet{zhu17}.
The three groups consist of 17, 17 and 16 events which have the source $I$-band magnitude 
$I_{\rm S} \leq 17.75$, $17.75 < I_{\rm S} \leq 19.40$ and $I_{\rm S} > 19.40$, respectively.
These boundaries of $I_{\rm S}$ are seleccted to make the number of events in each group nearly equal
so that results of the AD tests below are not dominated by the different sample sizes. The sensitivity
of these subsamples of 16 or 17 events is reduced compared to the full sample of 50 events.
To evaluate this effect, we have conducted Monte Carlo simulations of $10^4$ AD tests each with 
randomly selected 16 and 17 event sub-samples of the full Z17 sample. These yield median $p_{\rm AD}$
values of $p_{\rm AD} = 0.026$ and $p_{\rm AD} = 0.022$ for the 16 and 17 event
subsamples, respectively, which compare to $p_{\rm AD} = 5.1 \times 10^{-5}$ for the full 50 event sample.
Note that, although the source flux in IRAC 3.6 $\mu$m may seem to more directly represent the brightness of 
events seen by {\it Spitzer}, we do not choose it here because the source flux is one of the 
model parameters and measurements of the source flux in 
IRAC 3.6 $\mu$m could be contaminated by the systematic errors in {\it Spitzer} data. 
On the other hand, the source flux in $I$-band is well measured by OGLE data and is unlikely to be 
affected by the systematic errors.

Figure \ref{fig-3Is} shows plots similar to Figures \ref{fig-tEvtil}(c) and \ref{fig-B14_NF2}(c), but for each of the 
three source brightness sub-samples with no error bar inflation (i.e., $k = 1.0$).
In panels (a)-(c), the five curves are the median, $1\sigma$ and $2\sigma$ curves from the B14 model.
Panel (a) shows that the brightest stars still tend to have $\widetilde{ \pi}_{\rm E, post}$ values larger than the model
predictions, but the distribution of $\pi_{\rm E}$ values for the $17.75 < I_{\rm S} \leq 19.40$ subsample
(b) and especially the faint event subsample (c) are much more strongly skewed toward larger $\pi_{\rm E,obs}$
values. This is quantitatively confirmed by the $p$-values 
from AD-tests for the CDFs of the inverse percentile plotted in panel (d), where $p_{\rm AD} = 0.26$, $0.045$ and $3.4 \times 10^{-4}$ are obtained for the 
bright, middle and faintest subsamples, respectively. 
This result indicates that there is a correlation between $p_{\rm AD}$ values and the source 
brightness of events used in the test. Table \ref{tab-resultsIs} shows the results of these AD tests with
all three Galactic models.

\begin{deluxetable}{ccccc}
\tabletypesize{\normalsize}
\tablecaption{Results of AD tests on events in three different source brightness ranges. \label{tab-resultsIs}}
\tablehead{
\colhead{Model} & \multicolumn{3}{c}{$A^2$ ($p_{\rm AD}$) w/ $k=1$} & \colhead{$k\,(p_{\rm AD} = 0.05)$ for} \\
\cline{2-4}                   
\colhead{}                    & \colhead{$I_{\rm S} \leq 17.75$} & \colhead{$17.75 < I_{\rm S} \leq 19.40$}  & \colhead{$I_{\rm S} > 19.40$} & $I_{\rm S} > 17.75$ events\tablenotemark{a}\\
\colhead{}                    & \colhead{(17 events)}            & \colhead{(17 events)}                     & \colhead{(16 events)}         &  
}
\startdata
Z17  & 1.7 ($0.13$)                     & 3.1 ($0.026$)                & 5.7 ($1.4 \times 10^{-3}$)  & 2.77 \\
S11  & 1.2 ($0.27$)                     & 2.4 ($0.054$)                & 7.0 ($3.7 \times 10^{-4}$)  & 2.61 \\
B14  & 1.2 ($0.26$)                     & 2.6 ($0.045$)                & 7.0 ($3.4 \times 10^{-4}$)  & 2.61 \\
\enddata
\tablenotetext{a}{Minimum value of error inflation factor $k$ to be $p_{\rm AD} > 0.05$ but keeping $k = 1$ for events with $I_{\rm S} \leq 17.75$.}
\end{deluxetable}

This result is consistent with the idea that the systematic errors have a larger effect
on the parallax measurement for fainter sources. In particular, the 
effects of the systematic errors on the parallax measurements are much smaller for the 17 events with 
$I_{\rm S} \leq 17.75$  than for the other sub-samples with fainter source stars.
The AD tests on this bright source star sub-sample give $p_{\rm AD} = 0.13$, 0.27 and 0.26
for the Z17, S11 and B14 models, respectively.
Thus, they are formally consistent with all the three Galactic models used in this paper if we apply the 
acceptable $p$-value range of $p_{\rm AD} \geq 0.05$ that is used in Section \ref{sec-artest},
although part of the improvement compared to the full sample of 50 events is due to the statistical
noise for a smaller sample size.
Note that the selection bias due to the preference to include brighter source
star in the {\it Spitzer} sample inserts very little bias into the measured $\pi_{\rm E}$ distribution,
as discussed in Section \ref{sec-gammaDs}.

\subsubsection{Correlation with Light Curve Peak Coverage by {\it Spitzer}}  \label{sec-corrpeak}
The role of observed data points for light curve modeling is highly dependent on what part of the light 
curve they cover. To measure the microlensing parallax using {\it Spitzer} for a single lens event, the most important 
part is the light curve magnification peak because $\piEbold$ is given by 
\begin{equation}
\piEbold \simeq \frac{\rm AU}{D_{\perp}} \left(\frac{\Delta t_0}{t_{\rm E}}, \Delta u_0 \right) \ ,
\label{eq-piEspace}
\end{equation}
where $D_{\perp}$ is the separation between the Earth and {\it Spitzer} perpendicular to the event direction and $\Delta t_0$ and $\Delta u_0$ are differences of the time at the event peak, $t_0,$ and 
impact parameter, $u_0$, seen by the Earth from those seen by {\it Spitzer}.
Because $\piEbold \parallel \murelbold$, $\Delta t_0 \propto$ the component of $\piEbold$ in the direction of
the Earth-satellite separation and $\Delta u_0 \propto$ the perpendicular component of $\piEbold$. For the location
of the Galactic bulge microlensing fields at RA $\sim 18$h and DEC $\sim -30^\circ$, {\it Spitzer}
is largely to the West of Earth at the time of the {\it Spitzer} observations. Thus, $\Delta t_0$ largely determines
$\pi_{\rm E,E}$ and $\Delta u_0$ largely determines $\pi_{\rm E,N}$.
Because $t_0$ is time of peak magnification and $u_0$ is the impact parameter that determines the 
peak magnification, observations over the peak in a single lens light curve are crucial to definitively determine 
the microlens parallax. We also note that, in general, observations only near the light curve peak 
are not sufficient to determine $u_0$, and consequently $\pi_{\rm E, N}$.

\begin{figure}
\begin{minipage}{0.32\hsize}
\centering
\includegraphics[width=60mm]{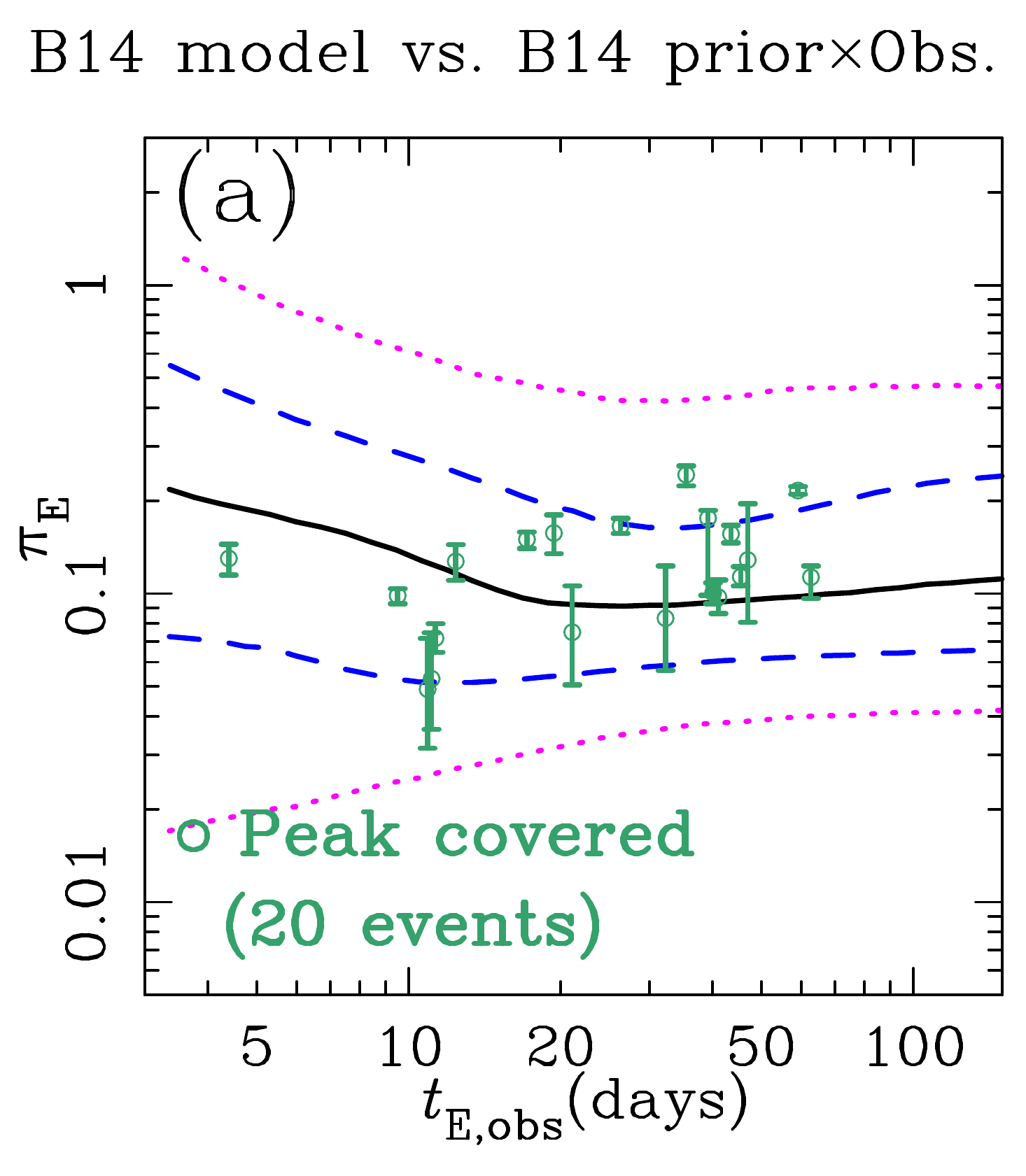}
\end{minipage}
\begin{minipage}{0.32\hsize}
\centering
\includegraphics[width=60mm]{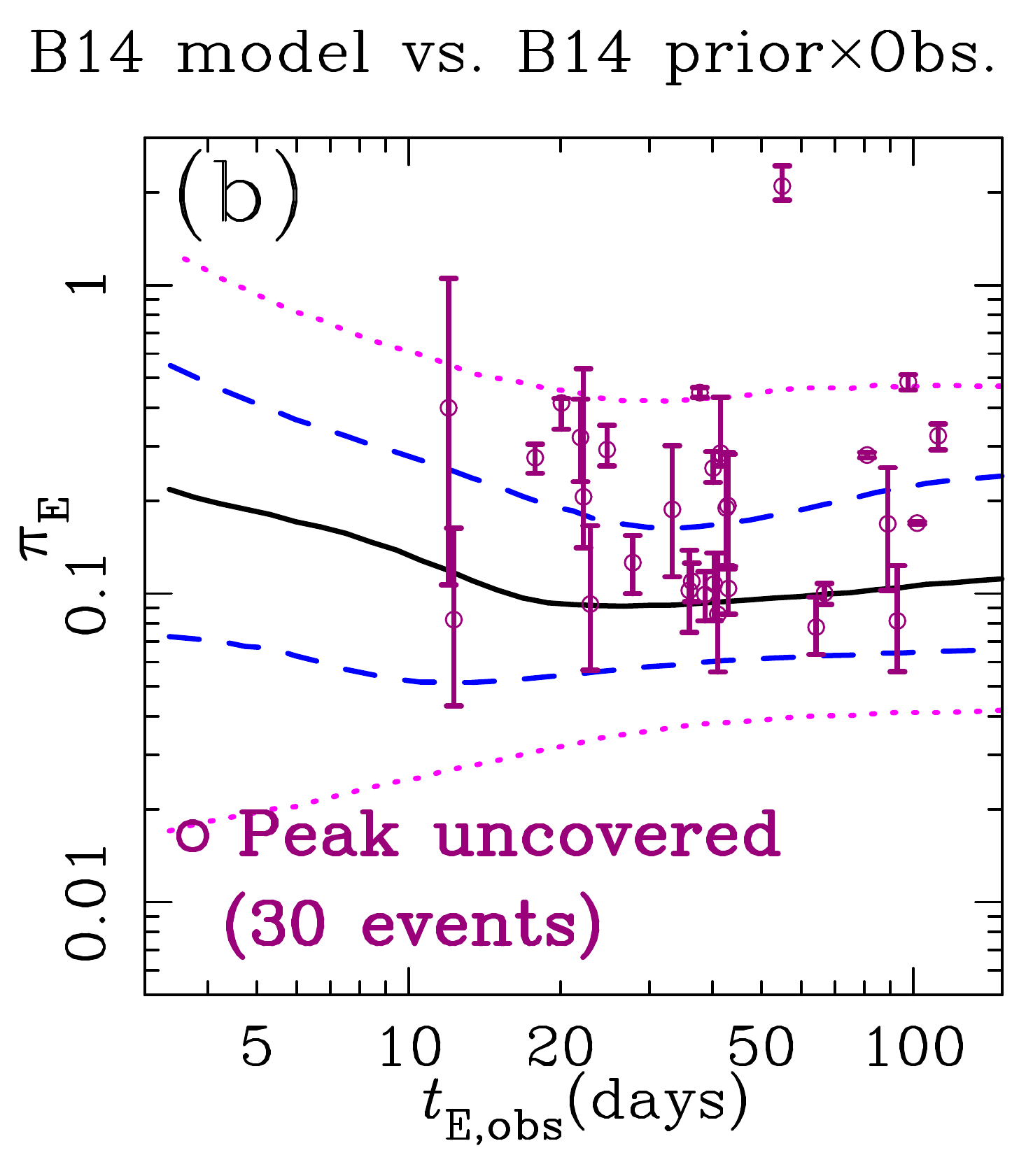}
\end{minipage}
\begin{minipage}{0.33\hsize}
\centering
\includegraphics[width=60mm]{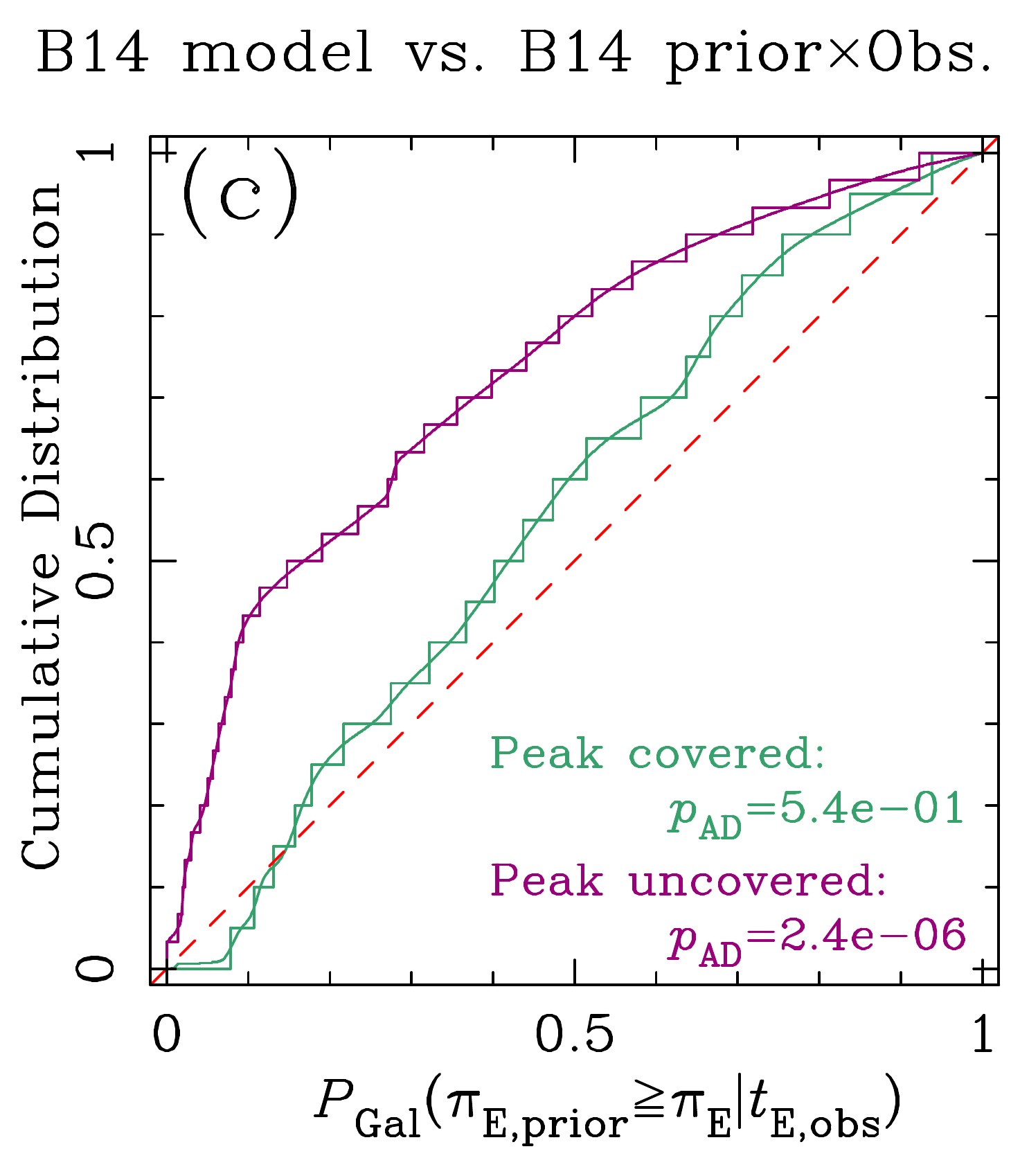}
\end{minipage}
\caption{Same as Figure \ref{fig-3Is}, but for the peak covered events and the peak uncovered events.}
\label{fig-peakpiE}
\end{figure}

The majority of the {\it Spitzer} events do not have the {\it Spitzer} observations over their peak. Instead, they
have data only for rising or declining part of the symmetric single lens light curve. Visual inspection of the
Z17 {\it Spitzer} data indicate that the $t_0$ value for the {\it Spitzer} can be easily determined for only 20 of 
the 50 Z17 events. These are events:
0379, 0565, 0798, 0843, 0958, 0961, 0965, 1161, 1172, 1188, 1189, 1256, 1341, 1370, 1420, 1448, 1450, 1457, 
1482 and 1533, where the numbers correspond to the event ``OGLE-2015-BLG-????".
Hereafter, we refer to these 20 events as the peak covered events and the other 30 events as the peak uncovered events.
Figures \ref{fig-peakpiE} (a)-(b) show the distributions of $t_{\rm E,obs}$ and $\widetilde{ \pi}_{\rm E, post}$ for these two 
samples and Figure \ref{fig-peakpiE} (c) shows results of AD tests on these samples using the B14 model.
These two samples have clearly different distributions from each other.
For the peak covered sample, we find the probability of $p_{\rm AD} = 0.54$ which is even larger than our 
fully acceptable threshold probability of $p_{\rm AD} = 0.50$ that is used in Section \ref{sec-test}. However, part of this improvement is 
due to the smaller sample size. Monte Carlo simulations of $10^4$ AD tests for a randomly selected 20 event
sub-sample of the Z17 sample give $p_{\rm AD} = 0.013$, which compares to 
$p_{\rm AD} = 5.1 \times 10^{-5}$ for the full 50 event sample.
For the peak uncovered sample, we find $p_{\rm AD}  = 2.4 \times 10^{-6}$.
This compares to $p_{\rm AD} = 2.0 \times 10^{-3}$ for a random sub-sample of 30 events and
$p_{\rm AD} = 5.1 \times 10^{-5}$ for the full sample. Obviously, the lack of {\it Spitzer} light curve peak coverage
leads to larger systematic errors.
Results with the S11 and Z17 models are shown in Table \ref{tab-resultspesks}, which confirm the same trend.

\begin{deluxetable}{rlrcrl}
\tabletypesize{\normalsize}
\tablecaption{Results of AD tests on events with and without peak coverage by {\it Spitzer}. \label{tab-resultspesks}}
\tablehead{
\colhead{Model} & \multicolumn{2}{c}{$A^2$ ($p_{\rm AD}$) w/ $k=1$} & \colhead{$k\,(p_{\rm AD} = 0.05)$ for}\\
\cline{2-3}
\colhead{}                  & \colhead{Peak covered\tablenotemark{a}} & \colhead{Peak uncovered\tablenotemark{b}} & peak uncovered events\tablenotemark{c}  \\
\colhead{}                  & \colhead{(20 events)}                   & \colhead{(30 events)}   & 
}
\startdata                                                                                                              
Z17   &  \ \ \ \  2.0 ($0.098$)                       & 9.4 ($2.6 \times 10^{-5}$)   & 2.94\\
S11   &  \ \ \ \  0.6 ($0.63$)                        & 11.6 ($2.7 \times 10^{-6}$)  & 2.56\\
B14   &  \ \ \ \  0.7 ($0.54$)                        & 11.7 ($2.4 \times 10^{-6}$)  & 2.61\\
\enddata
\tablenotetext{a}{Events that look like the peak position can be determined undoubtedly only from the {\it Spitzer} data. Details are seen in Section \ref{sec-corrpeak}.}
\tablenotetext{b}{Events other than the peak covered events.}
\tablenotetext{c}{Minimum value of error inflation factor $k$ to be $p_{\rm AD} > 0.05$ but keeping $k = 1$ for the 20 peak covered events.}
\end{deluxetable}

One issue in these two samples contain is a possible bias created due to the selection based on whether 
the {\it Spitzer} covered the event peak or not. Because the {\it Spitzer} observations start always 3 to 9 days 
after the event selection, it is somewhat easier for {\it Spitzer} to cover the event peak when it comes later 
than the peak seen by the Earth. However, the {\it Spitzer} event selection procedure helps to mitigate this
effect \citep{yee15} by preferentially selecting events that are alerted well before the peak.
This causes a slight bias in the $\pi_{\rm E, E}$ distribution between these two samples because $\pi_{\rm E, E}$ 
is mostly determined by $\Delta t_0/t_{\rm E}$ in Equation (\ref{eq-piEspace}), and this 
favors $\pi_{\rm E, E} < 0$ in the peak covered sample. However, the peak uncovered sample also has
a similar bias. Events with {\it Spitzer} observations only after the peak may have no real $\piEbold$ signal in
the {\it Spitzer}. If so, possible systematic errors would lead to an apparent $\pi_{\rm E, E} $ value less than
the true value. So, the $\pi_{\rm E}$ distribution could be biased because 
$\pi_{\rm E} = \sqrt{\pi_{\rm E, E}^2 + \pi_{\rm E, N}^2}$, but it is unclear if this bias would affect our tests.
We expect that whatever bias might exist in the selection of the peak covered and uncovered samples is
likely to be too small to explain the large difference in the $p_{\rm AD}$ values for these samples.

\subsubsection{Correlation with Obvious Photometric Errors in {\it Spitzer} Light Curves}  \label{sec-corrsys}
It is already known that many {\it Spitzer} light curves show obvious systematic errors 
\citep{pol16, shv17}, and \citet{zhu17} has a section that discusses these systematic errors. They pointed 
out five events where prominent deviations 
from the single lens model are seen and describe these are likely to be caused by systematic photometry
errors that could potentially affect the parallax measurements. 
We visually identify 19 events that have obvious systematic photometry errors with respect to the
single lens light curve models presented by \citet{zhu17}, including the five events identified by \citet{zhu17}.
The 19 events are 0081, 0379, 0461, 0529, 0565, 0703, 0958, 0961, 0965, 0987, 1096, 1188, 1189, 1204, 1297, 
1348, 1447, 1448, and 1492 where the numbers correspond to the event ``OGLE-2015-BLG-????".
Because we do not know how these obvious {\it Spitzer} photometry errors are caused, we are not sure 
what kind of selection bias is caused by this selection.
However, it seems very likely that these obvious systematic errors will be corrected with systematic errors
in the $\piEbold$ measurements.

\begin{deluxetable}{ccc}
\tabletypesize{\normalsize}
\tablecaption{Results of AD tests on events with and without obvious systematic errors in the {\it Spitzer} light curve. \label{tab-resultssys}}
\tablehead{
\colhead{Model} & \multicolumn{2}{c}{$A^2$ ($p_{\rm AD}$) w/ $k=1$}\\
\cline{2-3}
\colhead{}                  & \colhead{Obvious sys.\tablenotemark{a}}     & \colhead{No obvious sys.\tablenotemark{b}} \\
\colhead{}                  & \colhead{(19 events)} & \colhead{(31 events)}
}
\startdata
Z17  &  7.8 ($1.5 \times 10^{-4}$) &  3.0 ($0.029$)\\
S11  &  8.0 ($1.2 \times 10^{-4}$) &  2.6 ($0.046$)\\
B14  &  7.9 ($1.3 \times 10^{-4}$) &  2.8 ($0.035$)\\
\enddata
\tablenotetext{a}{Events with the {\it Spitzer} light curves that show obvious systematic photometry errors. Details are seen in Section \ref{sec-corrsys}.}
\tablenotetext{b}{Events other than the events with obvious systematic errors.}
\end{deluxetable}

Table \ref{tab-resultssys} shows results of AD tests on those two sub-samples.
There is a clear difference between them; $p_{\rm AD} = 1.3 \times 10^{-4}$ for the sub-sample with obvious 
systematic errors and $p_{\rm AD} = 0.035$ for the sub-sample without 
such obvious systematic errors, when compared with the B14 model. While these $p_{\rm AD}$ values differ
by two orders of magnitude, this is partly because the sub-sample with obvious systematic errors is
smaller, with 19 instead of 31 events. However, the $p_{\rm AD}$ value for the 31 event sub-sample without
obvious systematic errors is also unacceptably low. 
This is probably because some of the events have systematic errors in their $\pi_{\rm E}$ values, but 
the systematic errors are not obvious because they can be modeled by shifting the measured $\pi_{\rm E}$
value away from the correct value. 

\subsection{Other Curious {\it Spitzer} Parallax Measurements} \label{sec-coounter}

There are three published events with {\it Spitzer} parallax measurements have been interpreted
as implying that the lens systems that are located at $D_L=3$-$4\,$kpc with orbits
in the direction opposite of the disk rotation \citep{shv17,shv19,chu19}, or possibly perpendicular to
the disk rotation direction. All of these are caustic-crossing or caustic-approaching lenses with $\theta_{\rm E}$ measurements, 
so the lens system mass can be determined from equation~\ref{eq-mass}. Then, $\pi_{\rm rel}$ can be
determined from equation~\ref{eq-piE}. As a result of these additional constraints, the constraints on the
direction of the lens-source relative motion are much tighter than on the single lens events of \citet{zhu17}.
The prior probability of a lens in the disk orbiting in the counter-Galactic rotation direction for these three events is
1-$3\times 10^{-3}$ smaller than having the lens orbit in the direction of Galactic rotation. The probability
of having 3 such events in the sample of $\sim 20$ published {\it Spitzer} caustic-crossing events is 
no greater than $3\times 10^{-6}$. 
This probability seems too small even though it might be mitigated somewhat by selection effects and publication bias, as suggested by \citet{shv19}.
It is also possible our Galactic models may underestimate the prior probability for the counter-rotating stars,
although the B14 model does include thick disk and spheroid components. However, our disk models do not
capture the systematic variations in the mean rotational velocity in the inner disk \citep{gai18}.

In addition to the very unlikely prior probability, two of the three events, OGLE-2016-BLG-1195 \citep{shv17} 
and OGLE-2017-BLG-0896 \citep{shv19}, have neither a bright source star with $I_{\rm S} \leq 17.75$ 
nor {\it Spitzer} light curve that shows a clear peak feature. Thus, they share two of the features associated with
large $\pi_{\rm E,obs}$ systematic errors. The {\it Spitzer} light curve for OGLE-2016-BLG-1195 also has an 
apparent systematic error that is not fit by the light curve model. 
The other event, MOA-2016-BLG-231 \citep{chu19}, has a bright source star with $I_{\rm S} \sim 15.5$, but it has very poor coverage by {\it Spitzer} that starts from $\sim 25$ days after 
the $t_0$ of the event seen from the Earth, and the {\it Spitzer} light curve looks almost flat. So, these three 
events share 3, 2 and 1 of the features associated with systematic photometry errors, respectively.
Thus, it is very likely that one or more of these ``counter-rotating" events
has a spurious $\piEboldobs$ due to systematic errors in the {\it Spitzer} photometry.
We find reasonable kinematic solutions for two of the three events when we apply the Galactic prior and the error-bar inflation factor to them
while it is difficult for OGLE-2017-BLG-0896 even with $k = 3.4$. We recommend further investigation of the
{\it Spitzer} photometry for these events.

\subsection{Correcting {\it Spitzer} Microlensing Parallax Measurements}\label{sec-adde}

In this subsection, we consider a number of different ways to improve the {\it Spitzer} microlensing
parallax measurements.  The most direct method would be to remove systematic errors 
from the {\it Spitzer} photometry. Prompted, in part, by an early version of this paper and an
associated presentation in the annual Microlensing Conference in January, 2019, two groups \citep{gou20,dan20}
have demonstrated methods that can remove some of these systematic errors.
The {\it Spitzer} microlensing team \citep{gou20} reports that they have obtained
baseline data for almost all planetary events
from their 2014-2018 campaigns in their last 2019 season to help characterize the systematic photometry errors.
This baseline data was used to find and remove photometry contaminated by systematic errors due to the 
rotation of the telescope during the {\it Spitzer} observations of the KMT-2018-BLG-0029 event 
\citep{gou20},
 so this 2019 baseline data has proved very useful for one planetary events observed by {\it Spitzer}.
However, the baseline data were only taken for planetary microlensing events, and the 
{\it Spitzer} microlensing team's planned analysis method includes a comparison of planetary microlensing
events to microlensing events without planetary signals \citep{yee15}. In fact, the Z17 sample is
designed to be an example of this type of comparison sample. So, it seems that it may
be difficult to remove the systematic errors from this comparison sample, even if it is possible to 
remove the systematic errors from the planetary events with the help of the 2019 baseline data.

 \citet{dan20} have adapted the pixel level decorrelation method to {\it Spitzer} microlensing
survey photometry. This method  was originally developed to
study secondary eclipses in transiting planet systems \citep{dem15}. \citet{dan20} found 
that this method could improve the photometry by a factor of 1.5--6 for the two events that they analyzed.
This pixel level decorrelation and the baseline data seems like a promising approach, so it would
be useful to see it applied to a much larger sample of events.

The {\it Spitzer} microlensing team has also introduced a new procedure for determining {\it Spitzer-only}
microlensing parallax measurements, which can be compared to the ground-based-only microlensing
parallax measurements for consistency. This might be able to identify some inconsistencies, but it 
did not detect the systematic errors found in the KMT-2018-BLG-0029 {\it Spitzer} light curve, which were
revealed with the analysis of the 2019 baseline data.
However, ground-based microlensing parallax measurements are also subject to errors due to 
systematic photometry problems.

Thus, it is worthwhile to consider how we might deal with the {\it Spitzer} light curve data if the
systematic errors cannot be removed. We have addressed this in Section \ref{sec-test}, where we
adjusted the error bars for the measured microlensing parallax values.
We found that an error bar inflation factor $k = 2.2$ gives the measured $\pi_{\rm E}$ distributions that are
marginally consistent ($p_{\rm AD} = 0.05$) with the three Galactic models.
Given that our Galactic models have some simplistic features such as constant velocity dispersion regardless of Galactic distance
which is inconsistent with recent Gaia measurements \citep{gai18}, 
we consider that $k = 2.2$ is reasonable combined with modifications of such features in the Galactic model.
Because a reasonable error inflation factor might be different depending on the event characteristics, we 
conduct the same analysis for determining $k (p_{\rm AD} = 0.05)$ but keeping the factor $k = 1$ for the 17 bright source events ($I_{\rm S} \leq 17.75$).
This is based on a possible thought that the bright source events do not suffer from systematic errors and so $k > 1$ should be applied only to the other faint events.
We found a larger $k (p_{\rm AD} = 0.05)$ value of 2.61 for those faint events with the B14 model. The values for all the three models are shown in Table \ref{tab-resultsIs}.
We repeat this analysis to determine the $k (p_{\rm AD} = 0.05)$ values for the peak uncovered events where $k = 1$ is applied for the 20 peak covered events, and 
the results are shown in Table \ref{tab-resultspesks}.
One of these factors might be applied to {\it Spitzer} events that have no baseline data especially when its signal is suspicious, such as the source star is faint and/or the peak is not covered by {\it Spitzer}.

\subsection{Comparison to Previous Studies}\label{sec-comp}

Two previous studies have attempted rough statistical studies of {\it Spitzer} microlensing
parallax measurements using samples of 8 \citep{zang19}  and 13 \citep{sha19} events 
that have been selected out of a much larger sample for early publication.
Both samples are selected from previously published events rather than any statistical selection,
so they suffer from publication bias, as the first events selected for publication are less likely
to have apparent photometry problems than a statistical sample of events.
In contrast, the \citet{zhu17} sample is focused on the development of statistical methods for a study
of the Galactic distribution of planets. Therefore, these other studies are much less suited 
for statistical studies than the \citet{zhu17} sample. 

Furthermore,
the sample sizes of these studies are so small that they can only marginally probe the level of systematic
errors that we report in this paper. The probability (from the binomial formula) of 80\% each sample having
$\pi_{\rm E} >$ the median prediction of the Galactic model by random chance is 0.08 for 8 events and 
0.03 for 13 events. These compare to probability of $1.2\times 10^{-5}$ for the 50 events of the \citet{zhu17}
sample.

Hence, the analysis presented in this paper is the first statistical
analysis of a large unbiased sample of {\it Spitzer} microlensing parallax measurements.

\section{Summary and Conclusion} \label{sec-con}
We have compared the space parallax measurements of 50 single lens microlensing events from the 
2015 {\it Spitzer} microlensing campaign \citep{zhu17} with the predicted distribution from three
different Galactic models.
We found the following:

\begin{enumerate}
\item None of the three different Galactic models considered \citep{sum11,ben14,zhu17} can explain the 
observed distribution of measured $\pi_{\rm E}$ values. 
These $\pi_{\rm E}$ values are systematically 
larger than the predicted distribution even when the Galactic prior is applied, and Anderson-Darling test yields very low probabilities, 
$p_{\rm AD} \leq 6.6 \times 10^{-5}$, for all three models.
\item If we try to modify the Galactic models to restore consistency with the \citet{zhu17}  
parallax measurements,  we find that the disk to bulge mass ratio needs to be increased to be at least 3.1 times larger than 
the original value ($n_{\rm D/B} > 3.1$) for marginal consistency with the  Anderson-Darling test 
($p_{\rm AD} > 0.05$).
\item To be consistent with the recent studies of the stellar density in the solar neighborhood and the Galactic
bulge requires $n_{\rm D/B} = 1.68 \pm 0.12$ for the Z17 model. So, the value, $n_{\rm D/B} > 3.1$, required
for consistency with the \citet{zhu17} sample is clearly too large.
\item We find correlations between the $p$-values from the AD-tests and the following three characteristics of 
events in the sample used for our tests: the source magnitude, $I_{\rm S}$, 
the light curve peak coverage by {\it Spitzer}, and the presence of
obvious photometry errors seen in the {\it Spitzer} light curves.
\item We find that the systematic errors are substantially reduced for the 17 events with $I_{\rm S} \leq 17.75$ 
and the 20 events with the {\it Spitzer} coverage of the light curve peak. For these sub-samples, the 
excess of large $\pi_{\rm E}$ values measured is reduced to levels that are not statistically significant 
($p_{\rm AD} \geq 0.13$ and $p_{\rm AD} \geq 0.098$ for the bright and peak covered sub-samples, respectively).
However, some of this improvement is due to the small size of the sub-samples.
\item We find that the measured $\pi_{\rm E}$ distribution is still biased even for the events with no obvious systematic errors 
in the {\it Spitzer} light curve.
\item We find that the measured $\pi_{\rm E}$ distribution can be brought into marginal consistency ($p_{\rm AD} \geq 0.05$) with the
Galactic models by inflating the parallax parameter error bars by a factor of $k = 2.2$. 
This might be combined with a modest change of several simplistic structures in the Galactic models used to get a fully consistency.
\end{enumerate}

While the error bar inflation factor can bring the posterior $\pi_{\rm E}$ distributions into reasonable agreement with 
reasonable Galactic models, we believe that a more systematic investigation of the possible causes of a
these systematic errors is warranted, as discussed in Section \ref{sec-adde}. Our
investigation indicates that the systematic errors in the parallax measurements are significantly reduced for 
events with bright source stars. This suggests that there is a significant contribution to these systematic errors
from the blending of neighboring stars with the target stars. If so, methods developed to correct the photometry
for studies of transiting exoplanets are likely to be inadequate to correct the photometry for these
microlensing event targets.
We note that an obvious systematic trend in the {\it Spitzer} light curve is also confirmed in 
some bright events such as OGLE-2015-BLG-0448 \citep{pol16}, and \citet{dan20} shows that the 
pixel level decorrelation technique \citep{dem15} can improve the photometry.
Baseline photometry has already been collected in 2019 for most previously identified based,
in part, on an early version of this work \citep{gou20}, and this has been very helpful for at least some
of the events. Even if the 
sources of the systematic errors are not determined, it might be useful to characterize the error 
correlations so that significance of the $\piEboldobs$ measurements can be more accurately determined.
It seems clear that a better understanding of the systematic errors in the {\it Spitzer} photometry
is needed to realize the full scientific potential of the {\it Spitzer} microlensing program.

\acknowledgments
We thank Junichi Baba and Daisuke Suzuki for very helpful discussions.
Work by N.K. is supported by JSPS KAKENHI Grant Number JP18J00897.
D.P.B. was  supported by NASA through grant NASA-80NSSC18K0274.

\end{document}